\documentclass[%
 preprint, 
 superscriptaddress,
 amsmath,amssymb,
 aps, physrev,
]{revtex4-2}

\usepackage{graphicx}
\usepackage{dcolumn}
\usepackage{bm}

\begin{document}

\preprint{DOE-DEGREES}

\title{Theory of Transient Heat Conduction }

\author{David E. Crawford }
\thanks{These authors contributed equally to this work.}
\affiliation{Department of Physics, Auburn University, Auburn, Alabama 36849-5311, USA}

\author{Yi Zeng}
\thanks{These authors contributed equally to this work.}
\email[Correspondence email address: ]{yi.zeng@nrel.gov}
\affiliation{National Renewable Energy Laboratory, 15013 Denver West Parkway, Golden, CO 80401, USA}


\author{Judith Vidal}
\affiliation{National Renewable Energy Laboratory, 15013 Denver West Parkway, Golden, CO 80401, USA}%

\author{Jianjun Dong}
\email[Correspondence email address: ]{dongjia@auburn.edu}
\affiliation{Department of Physics, Auburn University, Auburn, Alabama 36849-5311, USA}%

\date{\today}

\begin{abstract}
Ultrafast and nanoscale heat conduction demands a unified theoretical framework that rigorously bridges macroscopic transport equations with microscopic material properties derived from statistical physics. Existing empirical generalizations of Fourier’s law often lack a solid microscopic foundation, failing to connect observed non-Fourier behavior with underlying atomic-scale mechanisms. In this work, we present a time-domain theory of transient heat conduction rooted in Zwanzig’s statistical theory of irreversible processes. Central to this framework is the time-domain transport function, $\overleftrightarrow{Z}(t)$, defined through equilibrium time-correlation functions of heat fluxes. This function generalizes the conventional concept of steady-state thermal conductivity, governing the transition of conduction dynamics from onset second sound type wave propagation at finite speeds to diffusion-dominated behavior across broad temporal and spatial scales. Unlike phonon hydrodynamic models that rely on mesoscopic constructs such as phonon drift velocity, our approach provides a quantitative and microscopic description of intrinsic memory effects in transient heat fluxes and applies universally to bulk materials at any temperature or length scale. By integrating atomistic-scale first-principles calculations with continuum-level macroscopic equations, this framework offers a robust foundation for numerical simulations of transient temperature fields. Furthermore, it facilitates the interpretation and design of transient thermal grating experiments using nanometer-scale heat sources and ultrafast laser systems in the extreme ultraviolet and x-ray wavelength ranges, advancing our understanding of heat dissipation dynamics.
\end{abstract}

\maketitle

\section{\label{sec:intro}Introduction}
The theory of heat conduction\cite{fourier1888theorie, carslaw1959conduction,  ziman2001electrons,  green1954markoff, callaway1959model,  zwanzig1965time, guyer1966thermal, majumdar1997microscale, chen2005nanoscale,dubi2011colloquium,li2012colloquium, biele2015time,Allen_Feldman_PhysRevB.48.12581, hua2019generalized,   chen2021non, simoncelli2019unified, simoncelli2020generalization, Zeng2021ITC, allen2022heat}, which describes a key mechanism for energy transfer in materials, is fundamental to basic physics and crucial for various technical applications, including microelectronics\cite{ majumdar1997microscale, chen2005nanoscale}, energy storage\cite{zeng2018molecular, zeng2021thermal}, and energy conversion technologies\cite{dong2001_PhysRevLett.86.2361,dubi2011colloquium}. Bridging critical gaps in the current theoretical framework of thermal transport processes is essential to address emerging challenges in ultrafast and nanoscale thermal phenomena\cite{cahill2003nanoscale, cahill2014nanoscale, koh2007_PhysRevB.76.075207, siemens2010quasi, minnich2011thermal,johnson2013direct, Hu2015SpectralMapping, hoogeboom2015mechanical}.

Conventionally, heat conduction is described at the continuum level by the heat diffusion equation (HDE)\cite{carslaw1959conduction}, which models the evolution of the temperature field, $T(\vec{r}, t)$, over time ($t$) and space ($\vec{r}$):
\begin{equation}
\label{eq:diffusion}
\frac{\partial T(\vec{r}, t)}{\partial t} - \vec{\nabla}_{r} \cdot \left( \overleftrightarrow{D} \cdot \vec{\nabla}_{r} T(\vec{r}, t) \right) = 0.
\end{equation}
Here, thermal diffusivity $\overleftrightarrow{D}$ quantifies a material’s capacity to conduct heat and is related to thermal conductivity $\overleftrightarrow{\kappa}$ through the relation $\overleftrightarrow{\kappa} = \rho C \overleftrightarrow{D}$, where $\rho$ is the mass density and $C$ is the specific heat capacity. The thermal conductivity, $\overleftrightarrow{\kappa}$, is defined by Fourier’s law of heat flux, $\vec{j}$, under \textit{steady-state} conditions:
\begin{equation}
\label{eq:fourier}
\vec{j}  = - \overleftrightarrow{\kappa} \cdot \vec{\nabla}_{r} T.
\end{equation}
Both heat conduction \textit{coefficients}, $\overleftrightarrow{D}$ or $\overleftrightarrow{\kappa}$,  can be measured experimentally or calculated through atomistic simulations\cite{ziman2001electrons}, providing essential inputs for modeling heat transfer in materials.

The adoption of Fourier’s law, however, introduces a theoretical limitation to the heat diffusion theory (Eq. \ref{eq:diffusion}), as it assumes a slowly varying, quasi-steady temperature field, $T(\vec{r}, t)$. While this assumption holds approximately for many macroscopic heat conduction processes, transient thermal measurements have consistently revealed significant deviations from Fourier’s predictions, particularly at very small length scales \cite{Frazer2019_PhysRevApplied.11.024042, 2019_EUV_TTG, hua2019generalized, beardo2021general, Jeong_etal_PhysRevLett.127.085901, mcbennett2023universal}. These deviations are further highlighted by experimentally observed wave-like heat transfer phenomena, such as second sound at low temperatures, which provide direct evidence of the breakdown of Fourier’s law. Such phenomena have been reported in superfluid Helium-II \cite{Peshkov1944}, crystalline materials \cite{ackerman1966second,jackson1970second,narayanamurti1972observation,koreeda2007second,beardo2021observation}, and low-dimensional systems \cite{huberman2019observation,ding2022observation}. These findings underscore the need for a theoretical framework that advances beyond Fourier’s law, offering a more accurate and cohesive description of heat conduction in these regimes.

Empirical extensions of Fourier’s law, applied beyond steady-state conditions or at confined length scales, have been extensively employed to explore macroscopic non-Fourier phenomena\cite{joseph1989heat}. Among these generalizations, the Maxwell-Cattaneo-Vernotte (MCV) model\cite{maxwell1866dynamical,cattaneo1948atti,vernotte1958paradoxes} is particularly prominent. It predicts damped heat waves by introducing a time derivative term for the heat currents, thereby accounting for the memory effect in transient heat conduction. However, such empirical models often depend on parameters with limited microscopic justification and are typically fitted to experimental data. Although phenomenologically useful, these models are conceptually constrained, limiting deeper insights into the underlying physics.

Rigorous atomistic-scale quantum mechanical treatments of heat transport in many-body material systems\cite{biele2015time} remain highly challenging at present. As a result, recent first-principles approaches have primarily focused on calculating the lattice thermal conductivity of perfect crystals. These efforts have been facilitated by advancements in computing harmonic phonon spectra\cite{PhysRevB.60.950,PhysRevB.74.014109,ward2009ab,tang2009pressure,phonopy,li2014shengbte,feng2016_PhysRevB.93.045202,ravi2020_PhysRevX.10.021063} and anharmonic phonon-phonon scattering rates\cite{Tang4539,tang2014thermal,phono3py}. Building on the success of first-principles calculations of steady-state thermal conductivity, researchers have turned their attention to deriving non-Fourier terms in phenomenological heat equations. These efforts aim to avoid dependence on empirical data fitting\cite{hua2014analytical,de2014thermal,lee2015hydrodynamic,cepellotti2015phonon,cepellotti2017transport,alvarez2018thermal,simoncelli2020generalization,chiloyan2021green,honarvar2021directional}.

Numerous efforts have been made within the framework of phonon hydrodynamics\cite{ward1952iii,callaway1959model,chester1963second,sussmann1963thermal,enz1968one,guyer1966solution,guyer1966thermal} to explore heat transport in a bulk phonon gas (PG) under the hypothesis that phonon quasi-momentum is nearly conserved when Normal phonon scattering processes dominate. One of the most prominent phonon hydrodynamic models is the Guyer-Krumhansl equation (GKE)\cite{guyer1966solution,guyer1966thermal}, which introduces temporal memory and spatially non-local terms based on the mesoscopic concept of phonon drift velocity field. Although the terms in the GKE and its variants can be linked to first-principles results\cite{lee2015hydrodynamic,cepellotti2015phonon,PhysRevB.105.165303}, phonon hydrodynamics remains semi-phenomenological, with foundational concepts often oversimplified or lacking rigorous statistical derivation and experimental validation. Notably, the memory effects in the GKE are mathematically identical to those in the MCV model\cite{guyer1966thermal}. A major limitation of the hydrodynamic framework is its reliance on the dominance of momentum-conserving Normal phonon-phonon scattering over mechanisms like Umklapp scattering, defect interactions, and grain-boundary scattering\cite{guyer1966solution,guyer1966thermal}. This confines its applicability to lattice heat conduction in cryogenic temperature regimes and idealized low-dimensional materials. Despite its intrinsic limitations, the phonon hydrodynamic framework has regained attention as one of the few continuum modeling approaches capable of incorporating first-principles heat conduction properties. Additionally, its predictions of phonon second sound, akin to those of the MCV model, qualitatively align with experimental observations, supporting its application in regimes where Fourier’s law fails.

This study presents a theoretical framework that extends classical transport coefficients, such as thermal conductivity, from steady-state conditions to time-domain transport functions, capturing heat dissipation dynamics across all temporal and spatial scales. Utilizing the time-dependent phonon Boltzmann transport equation (td-phBTE) and Zwanzig’s statistical theory of irreversible processes, the framework identifies time-correlation functions (TCFs) of fluctuating heat currents as fundamental transport functions for heat conduction, directly computable through atomistic simulations. It predicts the universal onset of wave dynamics in heat conduction and a seamless transition from wave-like to diffusive behavior in all materials. Additionally, it offers a first-principles interpretation of phonon second sound in natural silicon crystals at all temperatures, without relying on assumptions from phonon hydrodynamics. By providing a statistically grounded and unified approach, the framework addresses the limitations of phenomenological models and difficult-to-verify mesoscopic constructs. Furthermore, it offers accurate insights and robust interpretations of experimental results from transient thermal grating spectroscopy\cite{johnson2013direct,2019_EUV_TTG}, advancing the understanding and application of nanoscale thermal physics and engineering.

\section{\label{sec:TDHCF}Time-Domain Heat Conduction Transport Functions}
To address the limitations of Fourier’s law in non-quasi-steady-state regimes, we analyze the transient heat current, $\vec{j}(t)$, in a homogeneous bulk material. At this stage, boundary and interface effects are not explicitly considered. The bulk material is assumed to be infinitely large and subjected to a small, spatially uniform, and time-dependent temperature gradient, ${\vec{\nabla}_{r} T}(t)$.

\subsection{\label{subsec:theory}Phonon Heat Currents and Memory Effects}
We begin by analyzing a bulk phonon gas (PG) system in a periodic crystal, where phonon modes are indexed by $\alpha$, representing the wavevector $\vec{q}$ in reciprocal space and the phonon branch index b (i.e., $\alpha = (\vec{q}, b)$). Each phonon mode is characterized by its frequency $\omega_\alpha$, group velocity $\vec{v}_\alpha$, and equilibrium occupation number $n_0(\alpha)$ at a reference temperature $T_0$.

The td-phBTE governs the time evolution of the phonon occupation number $n(\alpha, t)$. The mathematical approximations used in this analysis, detailed in Appendix \ref{sec:linear}, are consistent with those applied under steady-state conditions\cite{ziman2001electrons,zeng_and_dong_2019_vibFPE}. In numerical simulations, integration over reciprocal $\vec{q}$-space is approximated by summation over discrete $\vec{q}$-points, resulting in a total of $N_\text{mode}$ phonon modes. The linearized td-phBTE is expressed as:
\begin{equation}
\label{eq_td-phBTE_alpha}
\begin{aligned}
 & \frac{\partial n(\alpha,t)}{\partial t} + \frac{ \hbar \omega_{\alpha} \vec{v}_{\alpha} \vec{\nabla}_r T(t)}{4  k_B T_0^2 \sinh^2 \left(\frac{\hbar \omega_{\alpha}}{2 k_B T_0}\right)} = \\
& - \sum_{\beta=1}^{N_{\text{mode}}} L_{\alpha \beta} \cdot \frac{\sinh \left(\frac{\hbar \omega_{\beta}}{2 k_B T_0}\right)}{\sinh \left(\frac{\hbar \omega_{\alpha}}{2 k_B T_0}\right)}  \big(n(\beta,t) - n_{0}(\beta)\big),
\end{aligned}
\end{equation}
where $k_B$ is the Boltzmann constant, $\hbar$ is the Plank constant, and the phonon scattering matrix $L$ is a symmetric, semi-positive definite square matrix of dimensions $N_{\text{mode}} \times N_{\text{mode}}$.

An essential step in solving Eq.~\ref{eq_td-phBTE_alpha} analytically (see Appendix \ref{sec:solution}) is the use of the eigenvalues $\gamma_{\lambda}$ and eigenvectors $\vec{u}_{\lambda}$ of the $L$ matrix, defined as: $\sum_{\beta=1}^{N_{\text{mode}}} L_{\alpha \beta} u_{\lambda \beta} = \gamma_{\lambda} u_{\lambda \alpha}$. The eigenvalues and eigenvectors form a complete, orthonormal basis, with the eigenmode index $\lambda = 1, \ldots, N_{\text{mode}}$\cite{zeng_and_dong_2019_vibFPE}. To simplify the solution, we define new variables $\eta_{\lambda}(t)$ as: $\eta_{\lambda}(t) \equiv \sum_{\alpha=1}^{N_{\text{mode}}} u_{\lambda \alpha} \cdot 2 \sinh \left(\frac{\hbar \omega_{\alpha}}{2 k_B T_0}\right) \cdot \left(n(\alpha,t) - n_{0}(\alpha)\right)$. Using this definition, $n(\alpha,t)$ can be expressed as:  $n(\alpha,t) = n_{0}(\alpha) + \sum_{\lambda = 1}^{N_{\text{mode}}} \frac{u_{\lambda \alpha} \eta_{\lambda}  (t)  }{2 \sinh \left(\frac{\hbar \omega_{\alpha}}{2 k_B T_0}\right) }$. The derived solution for $n(\alpha,t)$, as detailed in Appendix \ref{sec:solution}, is expressed as:
\begin{equation}
    \label{eq_solution_n_of_t}
    n(\alpha,t) = n_{0}(\alpha)  - \sum_{\lambda, \beta = 1}^{N_{\text{mode}}} \frac{u_{\lambda \alpha} u_{\lambda \beta} \hbar \omega_{\beta} \Vec{v}_{\beta} \cdot \int_{-\infty}^{t}dt' e^{-\gamma_{\lambda} (t-t')} \Vec{\nabla}_r T(t')}{4k_B T_{0}^2 \sinh \left(\frac{\hbar \omega_{\alpha}}{2 k_{B} T_{0}} \right)  \sinh \left(\frac{\hbar \omega_{\beta}}{2 k_{B} T_{0}} \right)}.
\end{equation}

Using Peierls’ phonon heat flux formula, $\vec{j}_{PG} = \frac{1}{V_{\text{cell}} \cdot N_{\text{mode}}}\sum_{\alpha} \left( n(\alpha,t) - n_{0}(\alpha) \right) \hbar \omega_{\alpha} \vec{v}_{\alpha}$, for a PG with unit-cell volume $V{\text{cell}}$ and total $N_{\text{mode}}$  phonon modes, we show that the transient heat flux $\vec{j}_{\text{PG}}(t)$ in a bulk PG is not directly proportional to the instantaneous temperature gradient $\vec{\nabla}_{r} T(t)$, as predicted by Fourier’s law. Instead, $\vec{j}_{\text{PG}}(t)$ is expressed as a convolution integral, where the time-domain heat conduction transport function (td-HCTF), $\overleftrightarrow{Z}_{\text{PG}}(t)$, describes the causal memory effects of the transient $\vec{\nabla}_{r} T(t)$ in the past:
\begin{subequations}
\label{eq_j_PG}
\begin{eqnarray}
    \vec{j}_{\text{PG}}(t) =- \int_{-\infty}^{t} dt' \overleftrightarrow{Z}_{\text{PG}}(t-t') \vec{\nabla}_{r} T(t'), \label{eq_j_a} \\
     \overleftrightarrow{Z}_{\text{PG}}(t) = \sum_{\lambda=1}^{N_{\text{mode}}} \overleftrightarrow{\zeta}_{\lambda 0} e^{-\gamma_{\lambda}t},   \label{eq_j_b} \\ 
    \overleftrightarrow{\zeta}_{\lambda 0} =\sum_{\alpha,\beta=1}^{N_{\text{mode}}}
    \frac{ u_{\lambda \alpha}  u_{\lambda \beta}  \hbar \omega_{\alpha}  \hbar \omega_{\beta}  \vec{v}_{\alpha} \vec{v}_{\beta}} { 4 N_{\text{mode}} V_{\text{cell}} k_{B} T_{0}^{2} \sinh(\frac{\hbar \omega_\alpha}{2 k_{B} T_{0}}) \sinh(\frac{\hbar \omega_\beta}{2 k_{B} T_{0}})},\label{eq_mode_z0} 
\end{eqnarray}
\end{subequations}
where $\overleftrightarrow{\zeta}_{\lambda 0}$ characterizes the weighted contribution of the $\lambda$-th eigenmode of phonon collision matrices (Eq. \ref{eq_j_b}) to $Z(t)$. 

Notably, in the steady-state limit, where the temperature gradient $\vec{\nabla}_{r} T(t)$ becomes time-independent, the general heat flux formula, Eqs. ~\ref{eq_j_a}, reduces to Fourier’s law: $
\vec{j}_{\text{PG}} = - \overleftrightarrow{\kappa}_{\text{PG}} \vec{\nabla}_{r} T$, where the thermal conductivity is given by:
\begin{equation}
\label{eq_GK}
\begin{aligned}
\overleftrightarrow{\kappa}_{\text{PG}} =& \int_{0}^{\infty} \overleftrightarrow{Z}_{\text{PG}}(t) dt \\
= &\sum_{\lambda =1}^{N_{\text{mode}}} \overleftrightarrow{\zeta}_{\lambda 0}/ \gamma_{\lambda}, 
\end{aligned}
\end{equation}
where each eigenmode of the $L$ matrix contributes to $\overleftrightarrow{\kappa}$ through $\overleftrightarrow{\zeta}_{\lambda 0} /\gamma_{\lambda}$.

Our analysis demonstrates that, beyond steady-state thermal conditions, a single value of $\overleftrightarrow{\kappa}$ (Eq.\ref{eq_GK}) is insufficient to fully characterize the material properties governing heat conduction. To accurately model heat conduction under transient thermal conditions, we introduce a time-domain transport function, $\overleftrightarrow{Z}_{\text{PG}}(t)$ (Eqs.\ref{eq_j_a}–\ref{eq_mode_z0}), rigorously derived from the td-phBTE. This transport function, central to our approach, offers an analytical representation that directly links microscopic phonon properties, such as frequency, group velocity, and eigenmodes of the phonon scattering matrix, to macroscopic heat transport properties (Eqs.~\ref{eq_j_b} and \ref{eq_mode_z0}). Unlike previous works, such as those by Chiloyan et al.\cite{chiloyan2021green} and Hua and Lindsay\cite{hua2020space}, which employed solutions of the td-phBTE for frequency-domain temperature responses to heating or current response to temperature gradients, our framework derives an analytical form for $\overleftrightarrow{Z}_{\text{PG}}(t)$ under time-dependent, spatially uniform temperature gradients in an infinitely large crystal. This analytical formulation not only reduces computational complexity of description of transient heat conduction properties, but also generalizes the classical concept of thermal transport coefficients into the time domain functions, providing a powerful framework for exploring heat dissipation dynamics across diverse temporal and spatial scales.

Our formalism reveals the microscopic origins of macroscopic memory effects, previously introduced in empirical models\cite{gurtin_and_pipkin_1968_kern,nunziato_1971_memory,alvarez2007memory}. These effects arise from the interplay between heat current relaxation dynamics, which restore non-equilibrium states to equilibrium, and transient temperature gradients, which continuously redefine the equilibrium state. In a PG, mode-specific relaxation rates $\gamma_{\lambda}$ determine the duration of these effects, where larger $\gamma_{\lambda}$ values correspond to shorter memory times for the non-equilibrium heat current $\vec{j}_{\text{PG}}(t)$. The smallest $\gamma_{\lambda}$ values govern the timescale of heat current relaxation dynamics, determining the onset of deviations from the Fourier diffusion equation. By integrating these insights into a unified framework, our approach offers a robust, microscopic method for analyzing transient heat conduction in a PG, extending beyond the quasi-steady-state regime.

\subsection{\label{subsec:time}Time-Domain Zwanzig Formalism}
To generalize the findings of memory effects in transient heat currents from a PG to other materials, we compare the predicted $\overleftrightarrow{Z}_{\text{PG}}(t)$ functions (Eqs.~\ref{eq_j_PG}) with results from our earlier study on fluctuation-dissipation dynamics in a PG, derived using the Fokker-Planck equation\cite{zeng_and_dong_2019_vibFPE}. This comparison demonstrates that the $\overleftrightarrow{Z}(t)$ is directly proportional to the equilibrium TCFs of heat currents, $\langle \vec{j}_{\text{PG}}(t) \vec{j}_{\text{PG}}(0) \rangle_{\text{eq}}$.

This relationship between $\overleftrightarrow{Z}_{PG}(t)$ and $\langle \vec{j}_{\text{PG}}(t) \vec{j}_{\text{PG}}(0) \rangle_{\text{eq}}$ corresponds to Zwanzig’s theory of irreversible processes\cite{zwanzig1961memory, zwanzig1965time}, which establishes a universal connection between time-domain memory kernel functions in transport equations, which link relevant variables with their corresponding thermodynamic forces, and the TCFs of these variables. For heat conduction, the relevant variables are bulk heat fluxes ($\vec{j}_{\text{bulk}}$), and the corresponding thermodynamic forces are temperature gradients ($\vec{\nabla}_{r}T$). In Zwanzig’s formalism, macroscopic heat conduction processes are inherently non-Markovian due to the statistical reduction of the full microscopic dynamics to the dynamics of $\vec{j}_{\text{bulk}}$. Using a projection operator technique to re-derive the Green-Kubo formula for thermal conductivity\cite{green1954markoff}, Zwanzig’s theory treats the remaining microscopic variables as \textit{irrelevant variables} and explicitly connects memory kernel functions to the causal influence of $\vec{\nabla}_{r}T$ on transient $\vec{j}_{\text{bulk}}$. The memory kernel functions in Zwanzig’s heat conduction equations are equivalent to the $\overleftrightarrow{Z}(t)$ in our PG analysis. Consequently, our definition of $\overleftrightarrow{Z}(t)$ can be generalized to other materials, where it is expressed in terms of the equilibrium TCFs of fluctuating $\vec{j}_{\text{bulk}}(t)$: 
\begin{equation}
\label{eq:TDCK}
\overleftrightarrow{Z}(t) =
    \begin{cases} 
    0 , & \text{for } t < 0, \\
    \frac{V}{k_B T^2} \langle \vec{j}_{\text{bulk}}(t) \vec{j}_{\text{bulk}}(0) \rangle_{\text{eq}}, & \text{for } t \ge 0.
    \end{cases}
\end{equation}

Using the time-domain Zwanzig formalism (Eq.~\ref{eq:TDCK}), the transport equations for transient heat flux can be expressed as:
\begin{equation}
    \vec{j}(\vec{r}, t) =  - \int_{-\infty}^{+\infty} dt' \overleftrightarrow{Z}(t-t') \vec{\nabla}_{r} T(\vec{r}, t'), \label{eq_j1}\\
\end{equation}

Zwanzig \cite{zwanzig1961memory,zwanzig1965frequency} pointed out that Eq.~\ref{eq_j1} allows memory effects in the time domain to be represented mathematically in terms of frequency-dependent transport coefficients, as follows:
\begin{equation}
    \vec{J}(\omega) =- \overleftrightarrow{\kappa}(\omega) \cdot  \vec{\nabla}_{r} T(\omega), \label{eq_fd_1}, 
\end{equation}
where $\vec{J}(\omega)  =  \int_{-\infty}^{\infty} { dt \vec{j}(t) e^{-i \omega t} }$ and $\vec{\nabla}_{r} T(\omega)  =  \int_{-\infty}^{\infty} { dt \vec{\nabla}_{r} T(t) e^{-i \omega t} }$ are Fourier transformation of transient heat fluxess and temperature gradients, respectively. The frequency-dependent thermal conductivity $\overleftrightarrow{\kappa}(\omega)$ is defined as:
\begin{equation}
    \overleftrightarrow{\kappa}(\omega)  =  \int_{-\infty}^{\infty} { dt \overleftrightarrow{Z}(t) e^{-i \omega t}} = \int_{0}^{\infty} { dt \overleftrightarrow{Z}(t) e^{-i \omega t}}. \label{eq_kappa_freq}
\end{equation}

Although frequency-domain and time-domain formalisms are mathematically equivalent in many cases, the time-domain formalism offers notable conceptual and computational advantages for \textit{transient heat conduction} over the frequency-domain approach, an aspect largely overlooked in prior applications of Zwanzig’s formalism for heat conduction.
Specifically, the definition of $\overleftrightarrow{\kappa}(\omega)$  (Eq.~\ref{eq_kappa_freq}) relies on the sufficient integrability of $\overleftrightarrow{Z}(t)$ to ensure numerical convergence of the Fourier transform, a condition that is not always satisfied. For instance, in the absence of Umklapp phonon-phonon scattering processes in an infinitely large crystal, certain eigenvalues of the phonon scattering matrix approach zero, causing $\overleftrightarrow{Z}(t)$ functions to become non-integrable.

Moreover, for transient heat conduction, the time dependence of temperature fields is often pulse-like rather than periodic, making the time-domain approach inherently more suitable for numerical simulations. This advantage becomes particularly evident in Eq.~\ref{eq_j_b}, which suggests that $\overleftrightarrow{Z}(t)$ functions for $t \ge 0$ can be approximated as:
\begin{equation}
\label{eq_mode_z_approx}
   \overleftrightarrow{Z}(t) \approx \sum_{\lambda = 1}^M \overleftrightarrow{\zeta}_{\lambda_0} e^{-\gamma_{\lambda}t}, \text{with}\overleftrightarrow{Z}(0) \approx \sum_{\lambda = 1}^M  \overleftrightarrow{\zeta}_{\lambda_0},  
\end{equation}
where the number of terms, $M$, is determined by the desired numerical accuracy. Using this approximation, we introduce a new concept of mode-specific heat flux, $\vec{j}_{\lambda}(\vec{r}, t)$, as:
\begin{equation}
\label{eq:mode_specific_j}
    \begin{split}
     \vec{j}_{\lambda}(\vec{r}, t) & \equiv -\int_{-\infty}^t dt' \overleftrightarrow{\zeta}_{\lambda_0}e^{-\gamma_{\lambda}(t - t')}\vec{\nabla}_{r} T(\vec{r}, t')   \\
    & = -\int_{0}^{\infty} dt'  \overleftrightarrow{\zeta}_{\lambda_0}e^{-\gamma_{\lambda}(t')}\vec{\nabla}_{r} T(\vec{r},t - t'). 
    \end{split}
\end{equation}

We can demonstrate the following relation for numerical simulations of the time evolution of $\vec{j}(\vec{r}, t)$, bypassing computationally intensive time convolution calculations:
\begin{subequations}
    \begin{eqnarray}
    & \vec{j}(\vec{r}, t) = \sum_{\lambda = 1}^{M}   \vec{j}_{\lambda}(\vec{r}, t), \label{eqs_j_lambda-a}\\
    &\frac{\partial}{\partial t}\vec{j}_{\lambda}(\vec{r}, t) + \gamma_{\lambda} \vec{j}_{\lambda}(\vec{r}, t) = - \overleftrightarrow{\zeta}_{\lambda 0} \vec{\nabla}_{r} T(\vec{r}, t), \label{eqs_j_lambda-b}
    \end{eqnarray}
\end{subequations}
where $\lambda=1, ..., \text{M}$. It also becomes evident that the widely adopted empirical MCV model is a limiting case where $\overleftrightarrow{Z}(t)$ is approximated by a single exponential function, i.e., $\text{M}=1$ and $\dot{\overleftrightarrow{Z}}(t) \approx -\overleftrightarrow{Z}(0) e^{-t/\tau}$.

In summary, the time-domain Zwanzig formalism extends our findings in phonon gas systems to any materials, offering an efficient framework for modeling transient heat conduction. By approximating $\overleftrightarrow{Z}(t)$ as a sum of exponential terms, this approach simplifies numerical simulations while accurately capturing heat current dynamics. The mode-specific heat flux concept further enhances efficiency by bypassing time-convolution integrals.

\subsection{\label{subsec:Si}Results of Silicon Crystals}
\begin{figure*}[h]
    \includegraphics[width=\textwidth]{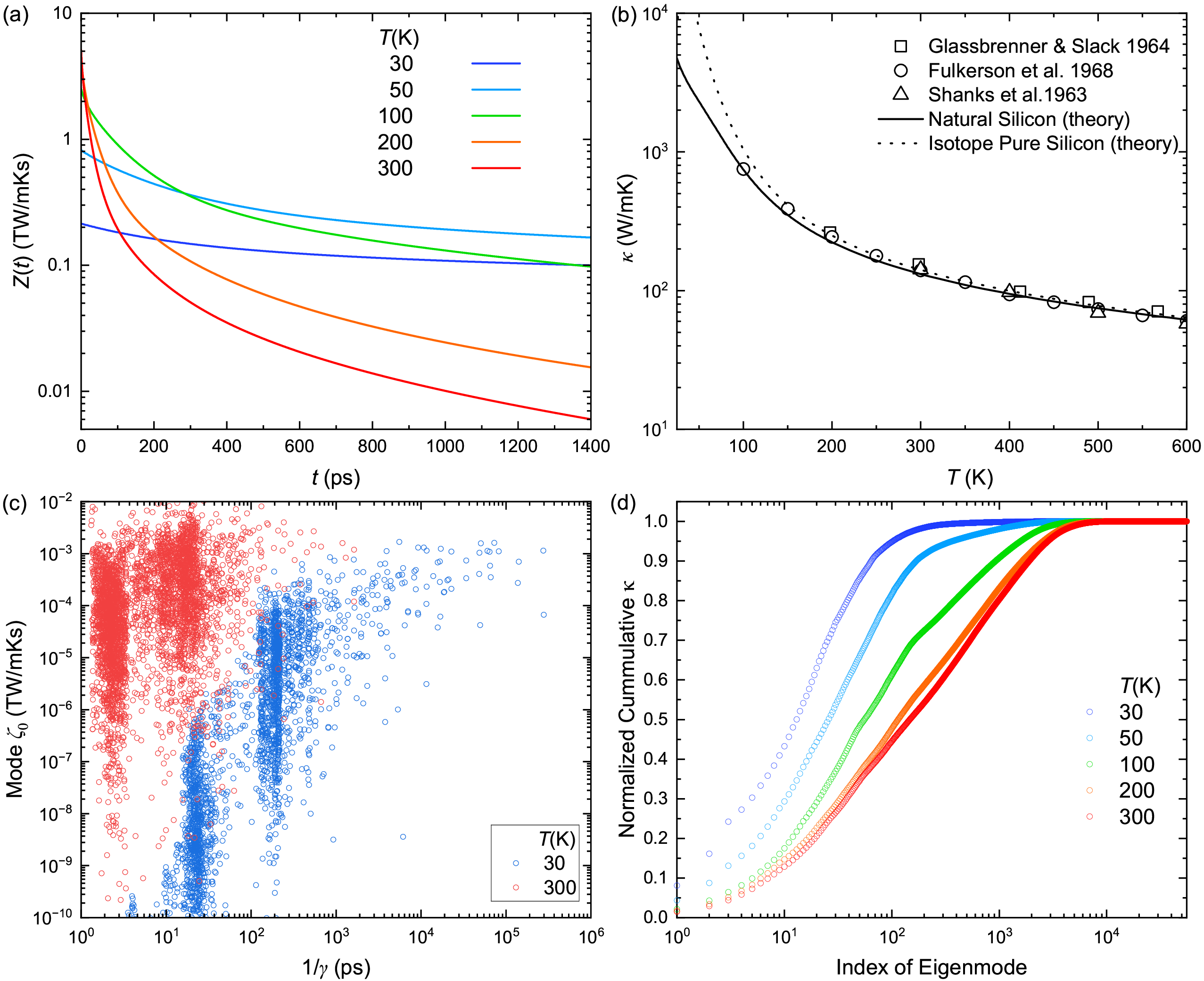}
    \caption{(a) First-principles calculated isotropic heat conduction transport functions $Z(t)$ for perfect natural Si crystals at 30 K, 50 K, 100 K, 200 K, and 300 K.  (b) First-principles calculated lattice thermal conductivity of isotope-pure (dotted line) and natural (solid line) silicon crystals from 25 K to 600 K. Available experimental data are shown as discrete symbols. (c) The individual mode contributions to the total $Z(t)$ in Si at 300 K (red circles) and 30 K (blue circles). Each circle represents one eigenmode, with its y-coordinate as the mode $\zeta_0$ and its x-coordinate as the mode decay time $1/\gamma$. (d) The relation between normalized cumulative thermal conductivity ($\kappa$) and the index of eigenmodes for natural silicon at various temperatures (30 K to 300 K). The eigenmodes are ranked based on their contributions to $\kappa$, and the curves highlight the disproportionate contributions of a small subset of eigenmodes.  } \label{fig:DFT_calculation}
\end{figure*}

We employ first-principles methods to calculate the $\overleftrightarrow{Z}(t)$ functions (Eqs.\ref{eq_j_b} and \ref{eq_mode_z0}) for natural silicon (Si) crystals, with calculation details provided in the Methods Section. {\bf Fig.\ref{fig:DFT_calculation}(a)} presents the temperature-dependent isotropic transport function, $Z(t) = \frac{1}{3} \left(Z_{xx} + Z_{yy} + Z_{zz} \right)$, derived from first-principles calculations for Si crystals at temperatures ranging from 30K to 300K. At 300K, $Z(t)$ decays rapidly, falling below 1$\%$ of its initial value $Z_0$ within 200ps. In contrast, at 30K, $Z(t)$ decays significantly slower, retaining nearly half of its initial value even at 1400~ps. These results illustrate the pronounced temperature dependence of $Z(t)$, with slower decay rates at lower temperatures.

Although $Z(t)$ itself has not yet been experimentally measured, its time integral (Eq.\ref{eq_GK}) corresponds to the steady-state thermal conductivity, $\kappa$, for which extensive experimental data are available. {\bf Fig.\ref{fig:DFT_calculation}(b)} compares our first-principles calculations of $\kappa$ as a function of temperature with experimental results\cite{PhysRev.130.1743, PhysRev.134.A1058, PhysRev.167.765}, demonstrating excellent agreement. Specifically, the calculated thermal conductivities ($\kappa_{\text{natural}}$) for natural silicon at 100K and 300K are 743W/mK and 132W/mK, respectively, closely matching the experimental values of 752W/mK and 140~W/mK reported by Fulkerson \textit{et al.}\cite{PhysRev.167.765}. This strong agreement validates our first-principles approach and underscores the reliability of the $Z(t)$-based framework for modeling heat transport.

The behavior of $Z(t)$ is determined by contributions from the eigenmodes of the bulk crystal's phonon scattering matrix, each contributing as $\zeta_0 e^{-\gamma t}$. Because the values of $\zeta_0$ and $\gamma$ differ by orders of magnitude across eigenmodes, the total $Z(t)$ does not exhibit a simple exponential decay, as often assumed in empirical models like MCV. {\bf Fig.\ref{fig:DFT_calculation}(c)} illustrates the distribution of eigenmode contributions, with each circle representing an eigenmode, its $y$-coordinate corresponding to $\zeta_0$ and $x$-coordinate to the decay time $1/\gamma$. The initial value, $Z(t=0)$, is the sum of all $\zeta_0$ contributions, while the overall decay time of $Z(t)$ is governed by the weighted average of $\gamma$ for eigenmodes with larger $\zeta_0$ values. Our calculations reveal that $Z(t)$ is overwhelmingly dominated by a small subset of eigenmodes with large $\zeta_0$ and long decay times ($1/\gamma$), particularly at lower temperatures. As the temperature decreases from 300K to 30K, the decay times of individual eigenmodes increase, and a few ultra-long decay modes with large $\zeta_0$ values become increasingly dominant. These collective modes, distinguished by their extended decay times and substantial contributions, align with phenomena described in phonon hydrodynamics by Guyer and Krumhansel\cite{guyer1966solution,guyer1966thermal} and are well-approximated by the kinetic collective model\cite{de2014thermal,PhysRevB.105.165303}. These findings emphasize the critical role of collective phonon dissipation mechanisms (Eq.~\ref{eq_mode_z0}), particularly at low temperatures, in shaping the decay behavior of $Z(t)$.

The disproportionate dominance of a small subset of eigenmodes in determining thermal conductivity is further highlighted in {\bf Fig.\ref{fig:DFT_calculation}(d)}. Each eigenmode contributes to $\kappa$ in proportion to $\zeta_0 / \gamma$. To evaluate the relative importance of these contributions, eigenmodes are ranked in descending order based on $\zeta_0 / \gamma$, with index 1 representing the largest contributor and $N_{\text{mode}}=55566$ the smallest. While all eigenmodes contribute to $\kappa$, the vast majority contribute negligibly. This observation enables significant computational efficiency when representing $Z(t)$ as a finite sum of $M$ exponential terms. By retaining only eigenmodes whose cumulative contributions exceed a specified threshold (e.g., 99$\%$ of $\kappa$), the number of modes required for computation can be drastically reduced without sacrificing accuracy. At lower temperatures, such as 30 K, fewer than 0.01$\%$ of modes account for approximately 30$\%$ of $\kappa$, reflecting the dominance of collective modes. At higher temperatures, the contributions are less concentrated, though the top 1$\%$ of modes still account for nearly 70$\%$ of $\kappa$. These results underscore the profound influence of a limited number of collective modes in governing thermal transport properties, particularly in the regime dominated by collective effects.


\section{\label{sec:unified_theory}Continuum Theory of Transient Heat Conduction}
\subsection{Governing Equations}
Our statistical analysis, grounded in Zwanzig’s theory of irreversible processes, has firmly established that the transient heat conduction properties of a homogeneous bulk material are quantitatively described by its time-domain heat conduction transport functions, $\overleftrightarrow{Z}(t)$ (Eq.~\ref{eq_j1}). These transport functions encapsulate the memory effects inherent in transient heat flux, providing a microscopic foundation for understanding non-Fourier heat conduction phenomena. Building upon this insight, and by integrating it with the fundamental principle of energy conservation, we derive a new governing equation for transient heat conduction in the following form:
\begin{equation}
\label{eq:zHCE}
\frac{\partial T(\vec{r}, t)}{\partial t} - \vec{\nabla}_{r} \cdot \frac{1}{\rho C } \big( \int_{0}^{+\infty} dt' \overleftrightarrow{Z}(t') \vec{\nabla}_{r} T(\vec{r}, t-t') \big)=0,
\end{equation}
where the term within $\big( \ldots \big)$ is equivalent to $\vec{j}(\vec{r},t)$ in Eq.\ref{eq_j1}, a result of the causality property of the $\overleftrightarrow{Z}(t)$ functions. This continuum equation, hereafter referred to as the Zwanzig heat conduction equation (zHCE), significantly extends the theoretical framework of heat conduction by addressing its limitations in ultrafast regime. Unlike the conventional Fourier HDE (Eq.\ref{eq:diffusion}), the zHCE incorporates the effects of finite relaxation times, enabling accurate modeling of transient heat conduction processes where Fourier HDE fails to apply.

The fundamental distinction between the zHCE and the Fourier HDE lies in the nature of the material-specific inputs required for their respective simulations. Numerical simulations based on the Fourier HDE (Eq.\ref{eq:diffusion}) depend on transport coefficients, such as thermal diffusivity, $\overleftrightarrow{D}$, or thermal conductivity, $\overleftrightarrow{\kappa}$, which are derived under steady-state or quasi-steady thermal conditions. In contrast, simulations employing the zHCE (Eq.\ref{eq:zHCE}) require time-domain heat conduction transport functions, $\overleftrightarrow{Z}(t)$, which rigorously account for causal memory effects and transient heat conduction dynamics. These $\overleftrightarrow{Z}(t)$ functions can be determined through atomistic simulations, including first-principles calculations, or experimentally constrained using data from transient thermal measurements. This key difference equips the zHCE with the capability to accurately model heat conduction phenomena that extend beyond the quasi-steady regime, addressing the limitations of the Fourier HDE in ultrafast thermal processes.

The Fourier HDE can be interpreted as a limiting case of the zHCE, applicable when the temporal variations in $T(\vec{r}, t)$ occur at rates significantly slower than the decay rates of $\overleftrightarrow{Z}(t)$.  In this quasi-steady regime, $\overleftrightarrow{Z}(t)$ decays to zero rapidly before any noticeable changes in $T(\vec{r}, t)$, resulting in $\vec{\nabla}_{r} T(\vec{r}, t-t’) \approx \vec{\nabla}_{r}T(\vec{r}, t)$ during the time intervals where $\overleftrightarrow{Z}(t’)$ is non-negligible. Effectively, the memory effects described by $\overleftrightarrow{Z}(t)$ are eliminated at the time scale of the slowly varying, quasi-steady temperature gradient $\vec{\nabla}_{r} T(\vec{r}, t)$. Under these conditions, the second term on the left-hand side of Eq.~\ref{eq:zHCE} reduces to an effective Fourier’s law:
\begin{equation}
\label{eq:quasi_steady}
\begin{split}
& \frac{1}{\rho C} \big( \int_{0}^{+\infty} dt’ \overleftrightarrow{Z}(t’) \cdot \vec{\nabla}{r} T(\vec{r}, t-t’) \big) \\
= & \frac{1}{\rho C}  \big( \int_{0}^{+\infty} dt’ \overleftrightarrow{Z}(t’) \big)  \cdot  \vec{\nabla}_{r} T(\vec{r}, t) \\
= & \overleftrightarrow{D} \cdot \vec{\nabla}_{r} T(\vec{r}, t),
\end{split}
\end{equation}
Consequently, the zHCE (Eq.\ref{eq:zHCE}) simplifies to the familiar form of the Fourier HDE (Eq.\ref{eq:diffusion}) in the quasi-steady regime, reaffirming that the conventional Fourier theory of heat conduction  is a limiting case of the more general Zwanzig theory of heat conduction under conditions of slowly varying temperature fields.

\subsection{Mode-Flux-Based Numerical Simulation Algorithm}
Direct computation of the time convolution integral in Eq.\ref{eq:zHCE} can be computationally prohibitive, especially for complex $\overleftrightarrow{Z}(t)$ functions or simulations involving large spatiotemporal domains. To address this challenge, we adopt the mode-specific approximation (Eqs.\ref{eq_mode_z_approx} and \ref{eq:mode_specific_j}), which reformulates the convolution integral into a mathematically equivalent system of coupled differential equations.

In this approach, a $\overleftrightarrow{Z}(t)$ function is expressed as a finite sum of $\text{M}$ exponential functions (Eq.\ref{eq_mode_z_approx}). The integer parameter $\text{M} \geq 1$ is chosen to balance numerical accuracy and computational efficiency. Substituting this approximation into Eq.\ref{eq:zHCE} transforms the original integral-differential equation into a system of $\text{M}+1$ coupled differential equations. These equations govern the evolution of the temperature field, $T(\vec{r}, t)$, and the mode-specific heat flux fields, $\vec{j}_{\lambda}(\vec{r}, t)$, where $\lambda = 1, \ldots, \text{M}$:
\begin{equation}
\label{eq:zHCE_mode}
\begin{cases}
    &\frac{\partial}{\partial t}T(\vec{r},t)= -\frac{1}{\rho C } \sum_{\lambda=1}^{\text{M}} \vec{\nabla}_{r}  \vec{j}_{\lambda}(\vec{r},t), \\
    &\frac{\partial}{\partial t}\vec{j}_{\lambda}(\vec{r}, t) = - \overleftrightarrow{\zeta}_{\lambda_0} \vec{\nabla}_{r} T(\vec{r}, t)- \gamma_{\lambda} \vec{j}_{\lambda}(\vec{r}, t). 
\end{cases}
\end{equation}

Eq.~\ref{eq:zHCE_mode} can be further discretized in both space and time to enable numerical simulations. For simplicity and to avoid unnecessary complexity in the mathematical details, we illustrate the algorithm in the one-dimensional case, represented as:
\begin{equation}
\label{eq:zHCE_delta}
\begin{cases}
    &T(x,t +\Delta t)= T(x,t)  -\frac{\Delta t}{2 \Delta x \rho C } \sum_{\lambda=1}^{\text{M}} \big(j_{\lambda}(x+\Delta x,t) - j_{\lambda}(x - \Delta x,t) \big), \\
    &j_{\lambda}(x, t+\Delta t) =  - \frac{\zeta_{\lambda_0} \Delta t}{2 \Delta x} T(x+\Delta x, t)  + \frac{ \zeta_{\lambda_0} \Delta t}{2 \Delta x} T(x,t) + (1 -  \gamma_{\lambda} \Delta t)   j_{\lambda}(x, t).
\end{cases}
\end{equation}
Here, $\Delta x$ and $\Delta t$ denote the spatial and temporal discretization steps, respectively, which are chosen to balance computational efficiency with numerical accuracy. By iteratively applying the finite difference scheme, the time evolution of  $T(x,t)$  and  $j_{\lambda}(x,t)$  can be numerically computed for a sufficiently small time step  $\Delta x$ and $\Delta t$ under specific initial condition at $t=0$.

As an illustrative example, we consider the time evolution of a one-dimensional temperature field in response to a spatially periodic heating pulse. The heating source is modeled as $h_{s}(x,t)=\rho C e^{iq_{0}x} \cdot \delta(t)$,  where the spatial wave number  $q_{0}$  defines the periodicity of the heating source with length scale  $d$, such that  $q_{0} = 2\pi/d$. For $t < 0$, the system is assumed to be in thermal equilibrium, uniformly maintained at a temperature  $T_{0}$ . At  $t = 0$, the application of the spatially periodic heating pulse imposes an initial unit temperature profile given by: $T(x,0) = T_{0} + e^{iq_{0}x}$, with the initial condition for all mode-specific heat flux components as $ j_{\lambda} = 0$ for all $\lambda$.

To solve the coupled governing equations of transient heat conduction (Eq. \ref{eq:zHCE_mode}), $T(x,t)$ and mode-specific $j_{\lambda}(x,t)$ are expressed  as $T(x,t) = T_{0} + e^{iq_{0}x} \cdot \Delta T_{q_{0}}(t), \quad \text{and} \quad j_{\lambda} (x,t) = \rho C e^{iq_{0}x} \cdot J_{q_{0},\lambda}(t)$, where  $\Delta T_{q_{0}}(t)$  represents the amplitude of the temperature field variation, and  $J_{q_{0},\lambda}(t)$  denotes the mode-specific heat flux amplitudes. For  $t > 0$ , the governing equations for the time evolution of  $\Delta T_{q_{0}}(t)$  and  $J_{q_{0},\lambda}(t)$  are discretized using the finite difference method, yielding:
\begin{equation}
\label{eq:TTG_delta}
 \begin{cases}
     &  \Delta T_{q_{0}}(t + \delta t) \approx  \Delta T_{q_{0}}(t) - i q_{0} \cdot \delta t \cdot \sum_{\lambda=1}^{\text{M}} J_{q_{0},\lambda}(t), \\
    & J_{q_{0},\lambda}(t +\delta t ) \approx   - i q_{0} \zeta_{\lambda_0} \delta t \cdot \Delta T_{q_{0}}(t)  + (1 -  \gamma_{\lambda} \delta t ) \cdot  J_{q_{0},\lambda}(t).
\end{cases}   
\end{equation}

While Eqs. \ref{eq:zHCE_delta} and \ref{eq:TTG_delta} are presented in one dimension for clarity, the generalization of these equations  to three-dimensional simulations is straightforward and involves extending the spatial derivatives and flux terms to account for the additional dimensions.

The newly proposed continuum governing equations (Eqs.\ref{eq:zHCE_delta} and \ref{eq:TTG_delta}) can also be reformulated into the framework of a generalized heat telegrapher heat equation (gHTE). This reformulation illustrates the diffusion-to-wave transitions in transient heat conduction phenomena. The detailed mathematical derivation of this gHTE, along with a thorough analysis of its connection to the underlying diffusion-to-wave transitions, is provided in Appendix \ref{sec:gHTE}.

\section{\label{sec:transient} Nanoscale Heat Conduction}
The Zwanzig theory of transient heat conduction offers a universal framework for modeling heat transport in homogeneous bulk materials. Although the current formulation does not explicitly incorporate boundary or interface effects, it establishes a rigorous theoretical foundation for understanding nanoscale heat conduction phenomena. Such phenomena frequently emerge in systems with highly spatially confined heating sources or sharply varying initial temperature fields,  $T(\vec{r})$. The general Zwanzig theory inherently links the length-scale features of nanometer spatial confinement to picosecond-scale transient heat conduction dynamics. These fast timescale dynamics, previously neglected in the conventional Fourier theory of heat conduction, are captured even in the absence of explicit boundaries or interfaces.

To explore the distinctive characteristics of transient heat conduction at nanoscale, we perform numerical simulations on silicon crystals with initial temperature profiles configured as either spatially localized or spatially periodic. For simplicity and to highlight the underlying physics, we focus on a one-dimensional case in this study. These simulations are based on Eqs.\ref{eq:zHCE_delta} and \ref{eq:TTG_delta}, with details of the parameters and computational setup provided in Section \ref{sec:method}. The temperature-dependent time-domain heat conduction transport functions, $\overleftrightarrow{Z}(t)$, and thermal conductivity, $\overleftrightarrow{\kappa}$, for silicon were derived from first-principles calculations (see Fig.~\ref{fig:DFT_calculation}).


\subsection{Universal Wave Dynamics}
\begin{figure}[ht!]
    \includegraphics[width=0.5\textwidth]{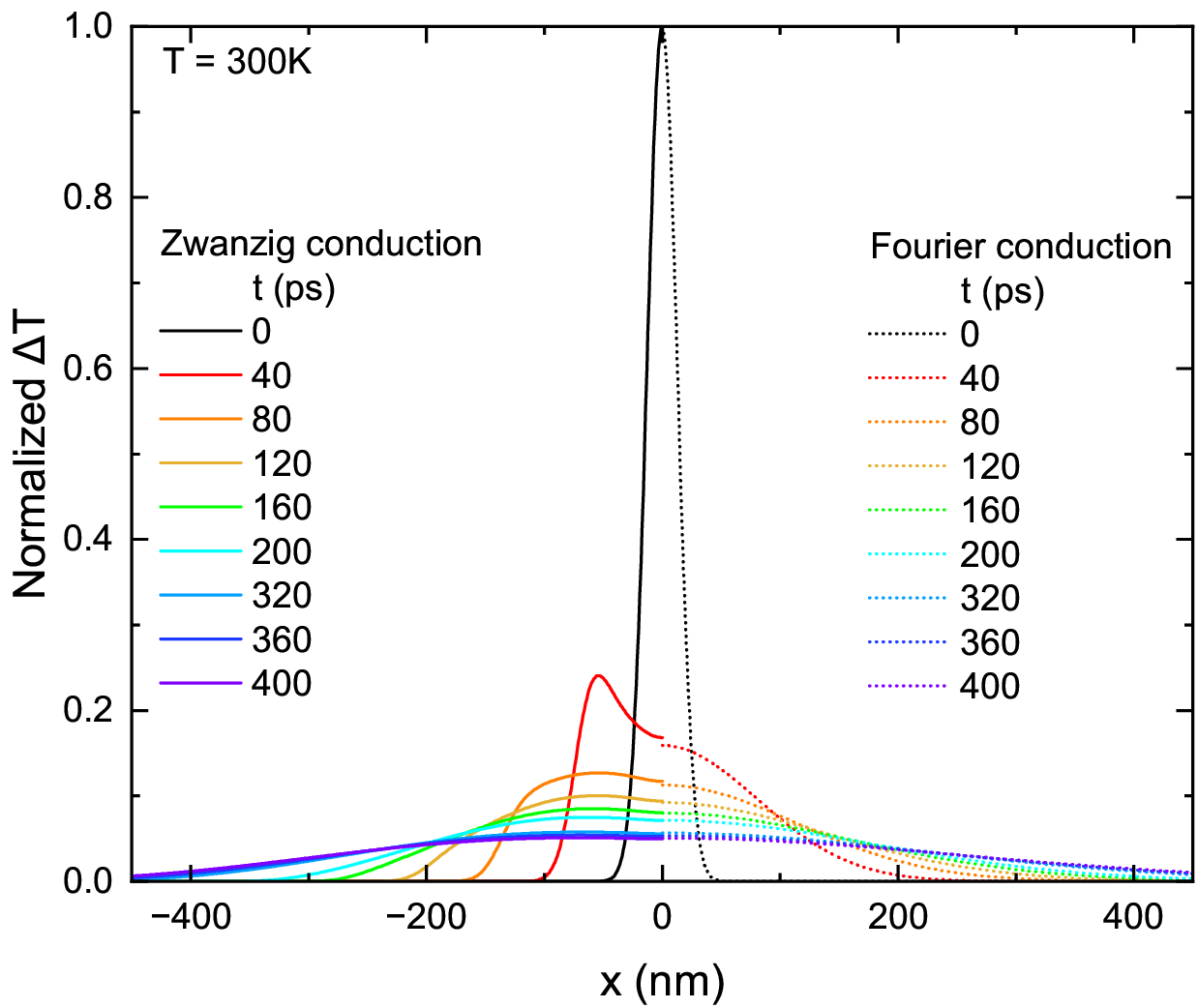}
    \caption{The side-by-side comparisons of the predicted temperature profiles between Zwanzig and Fourier theories at 300K. A more localized initial temperature profile is chosen at 300K, with a standard deviation ($\sigma$) of 12.8nm and confined within a full width of 128nm. A closer examination reveals fast-decaying wave-like features in heat conduction over tens of picoseconds at 300K, while Zwanzig theory predicts dominant wave dynamics lasting over nanoseconds at 30K. The plotted temperatures are normalized to 1 at the peak center of the initial profile.}
    \label{fig:Z_vs_F}
\end{figure}

To illustrate the distinctions between the Zwanzig and Fourier theories, we simulate the time evolution of one-dimensional temperature profiles in Silicon where the initial temperature is uniform except within a nanometer-sized region. Fig.~\ref{fig:Z_vs_F} provides a side-by-side comparison of the temperature profiles predicted by the two theories for silicon crystals.  Initially, the silicon crystal is at a uniform temperature of 300K, with an additional localized hot spot modeled as a Gaussian temperature profile with a standard deviation ($\sigma$) of 12.8nm, confined within a 128nm region. This setup represents a confined nanoscale heating scenario.  For silicon, the thermal diffusivity is $D_{\text{Si}} = 80~\text{mm}^2/\text{s}$ at 300~K, a regime where the Fourier diffusion theory is generally considered valid.

However, the comparison presented in {\bf Fig.\ref{fig:Z_vs_F}} reveals significant differences between the two theories within the first 100~ps timescales. Within the first 40~ps, the Zwanzig theory predicts wave-like heat conduction, characterized by the preservation of the shape of the initial Gaussian temperature profile. In contrast, the Fourier theory immediately predicts the diffusive spreading of the Gaussian temperature profile, consistent with its hallmark behavior of instantaneous diffusion. This comparison highlights the critical role of memory effects and transient dynamics, which are accurately captured by the Zwanzig framework but omitted in the Fourier description. These results highlight the critical importance of the Zwanzig theory in accurately describing ultrafast and nanoscale heat conduction phenomena, even in temperature regimes where the Fourier theory is conventionally deemed applicable.

As time progresses beyond 100~ps, the differences between the two predictions diminish rapidly. By approximately 120~ps, the Zwanzig theory transitions to a diffusion-like regime, converging with the predictions of the Fourier theory. At the long-time limit, both theories predict identical temperature evolution, as the Zwanzig theory reduces to the Fourier description in Silicon at 300~K.

These results demonstrate that while the Fourier diffusion theory provides an adequate description of long-time heat conduction in many materials, the Zwanzig theory is indispensable for accurately modeling transient heat conduction phenomena. This is especially critical for nanoscale systems and ultrafast time scales, where memory effects and wave-like dynamics significantly influence the heat transport process and are not captured by the Fourier framework.

\begin{figure*}[ht!]
    \begin{minipage}{0.3\linewidth} 
        \includegraphics[width=\textwidth]{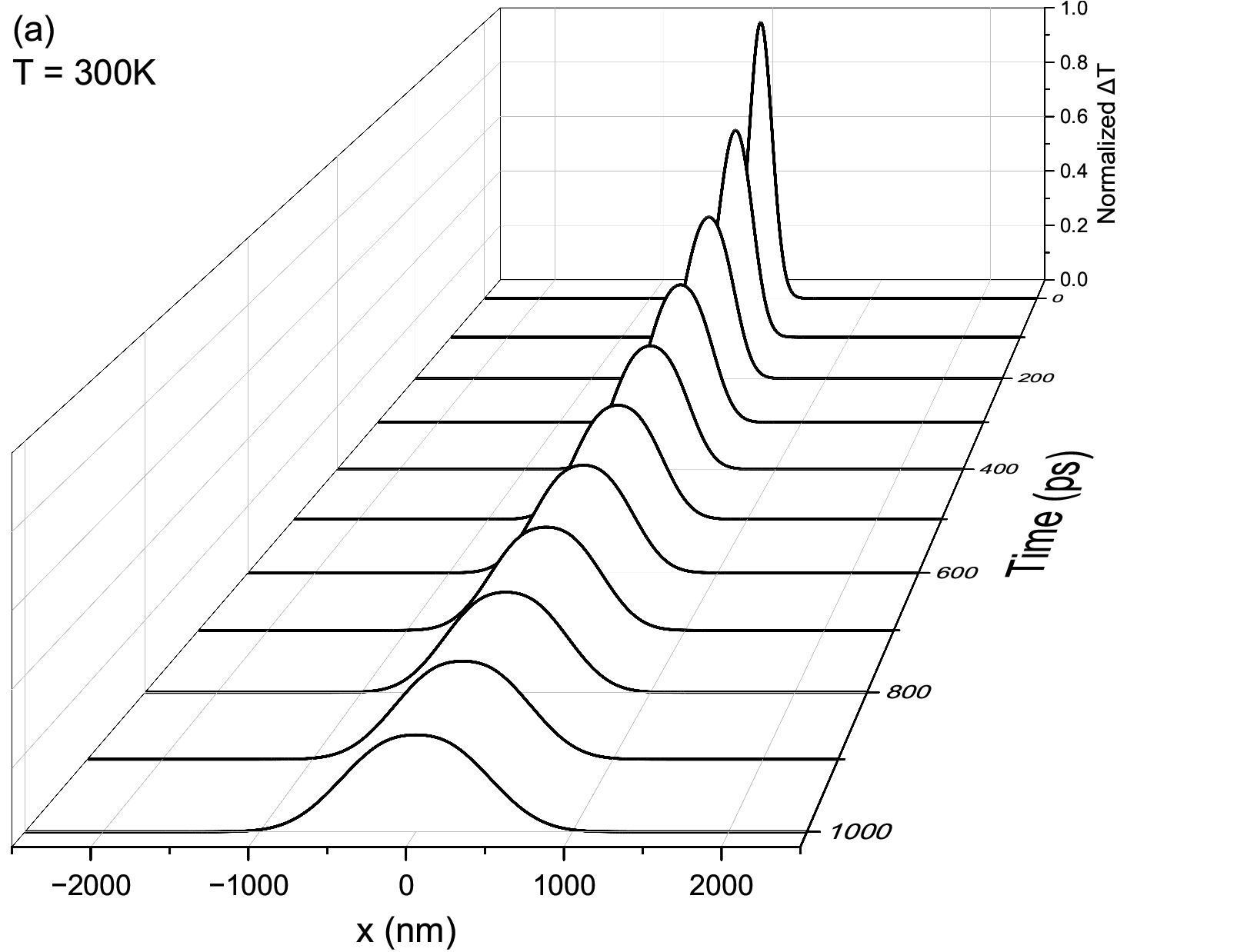}
    \end{minipage}
    \begin{minipage}{0.3\linewidth} 
        \includegraphics[width=\textwidth]{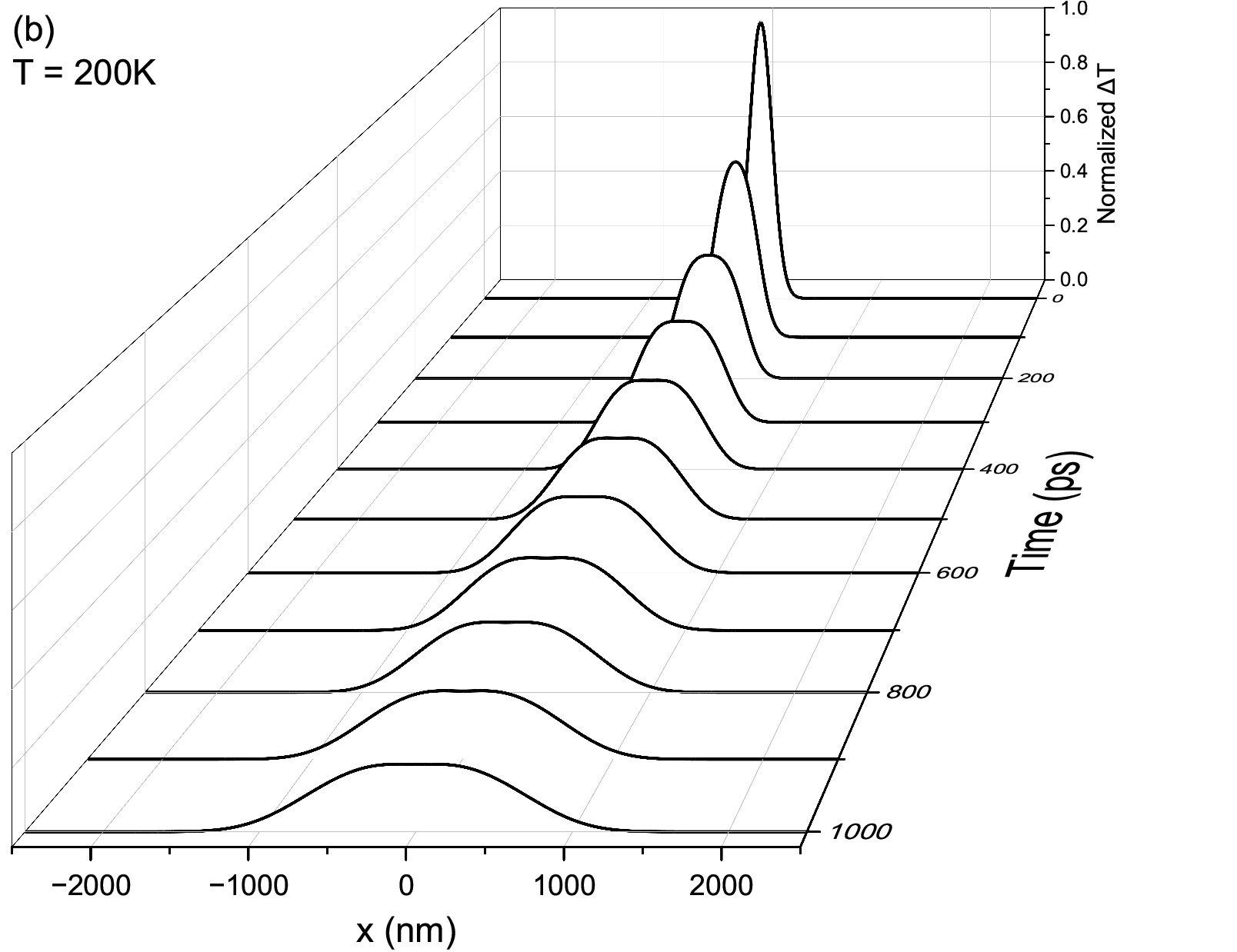}
    \end{minipage}
    \begin{minipage}{0.3\linewidth} 
        \includegraphics[width=\textwidth]{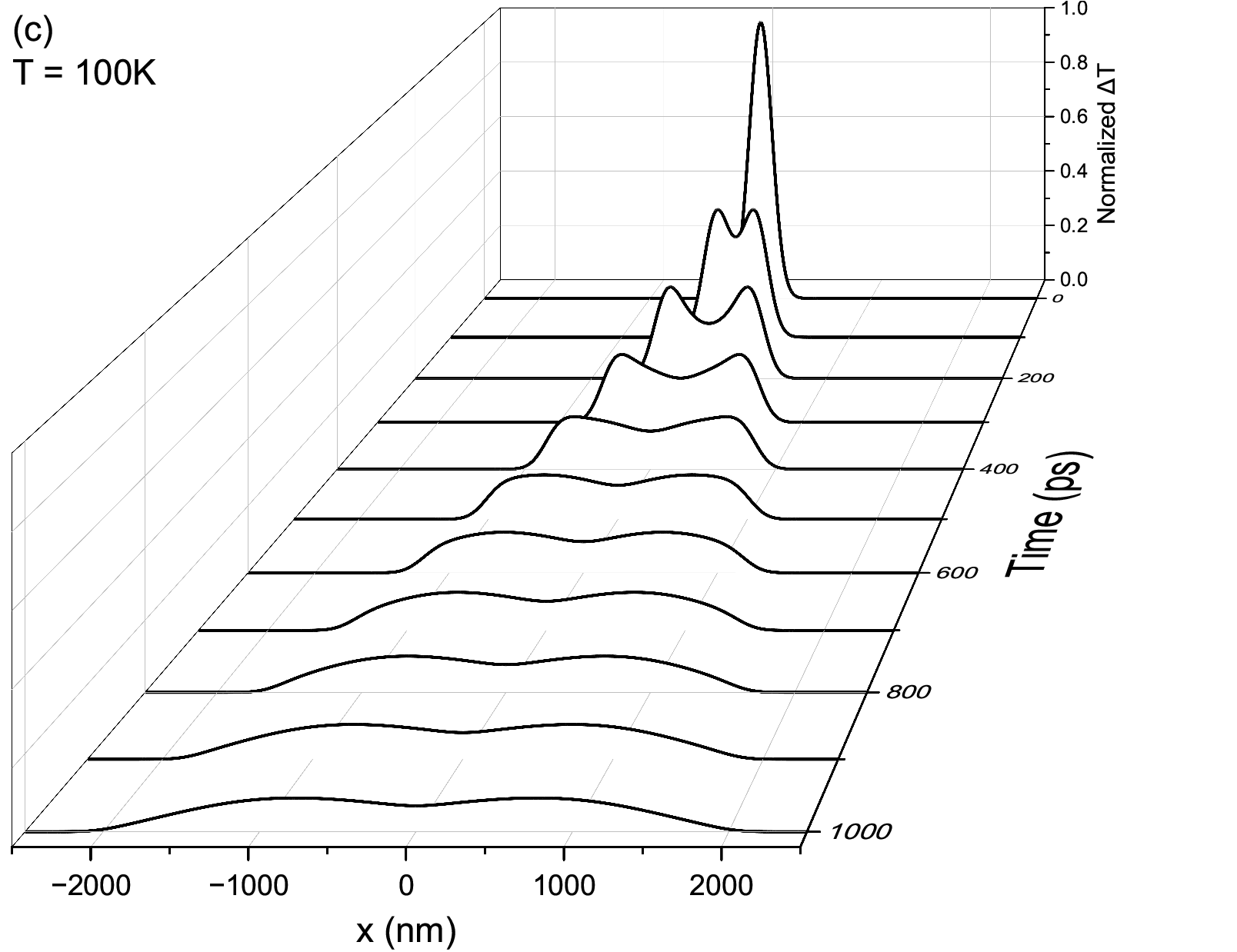}
    \end{minipage}
    \begin{minipage}{0.3\linewidth} 
        \includegraphics[width=\textwidth]{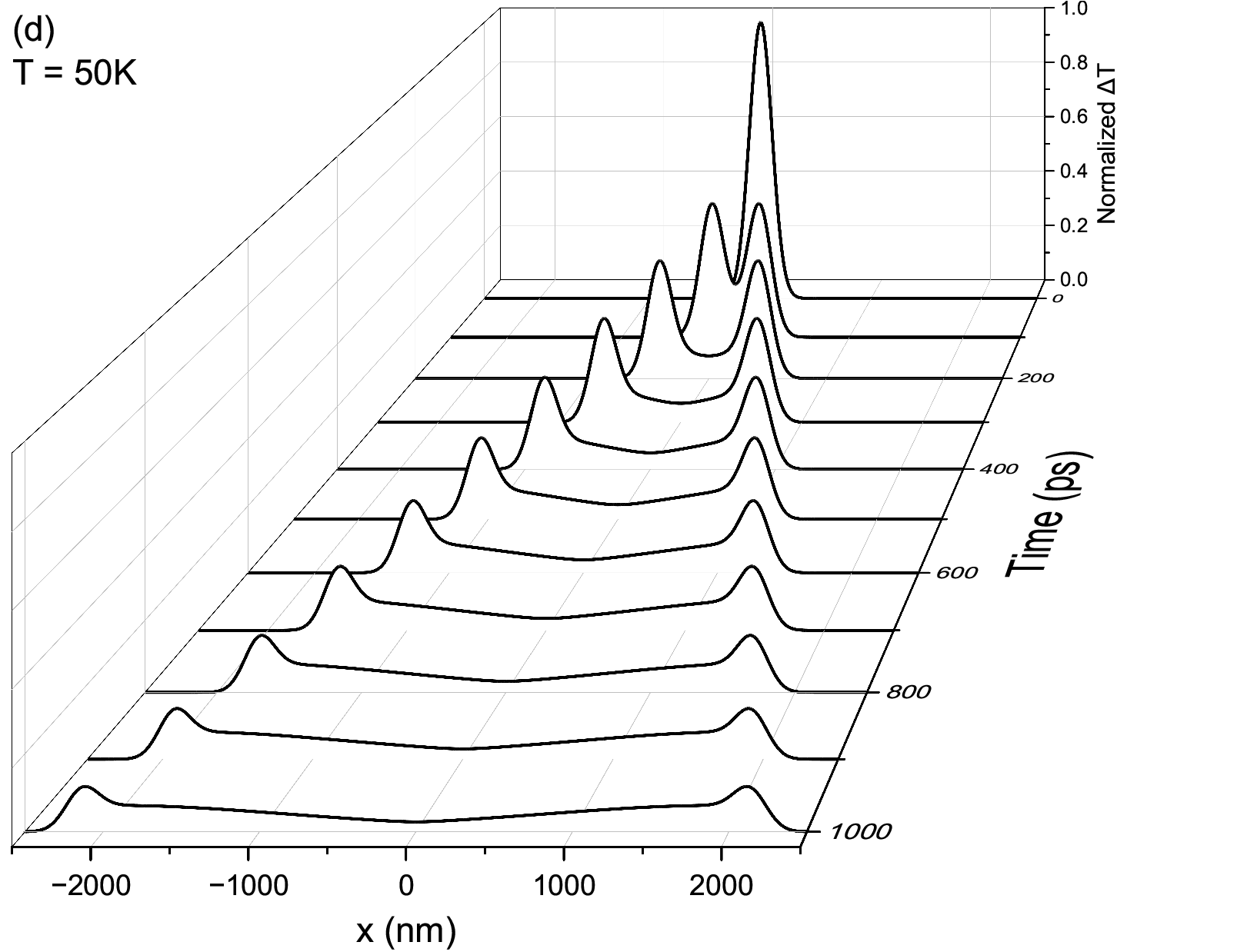}
    \end{minipage}
    \begin{minipage}{0.3\linewidth} 
        \includegraphics[width=\textwidth]{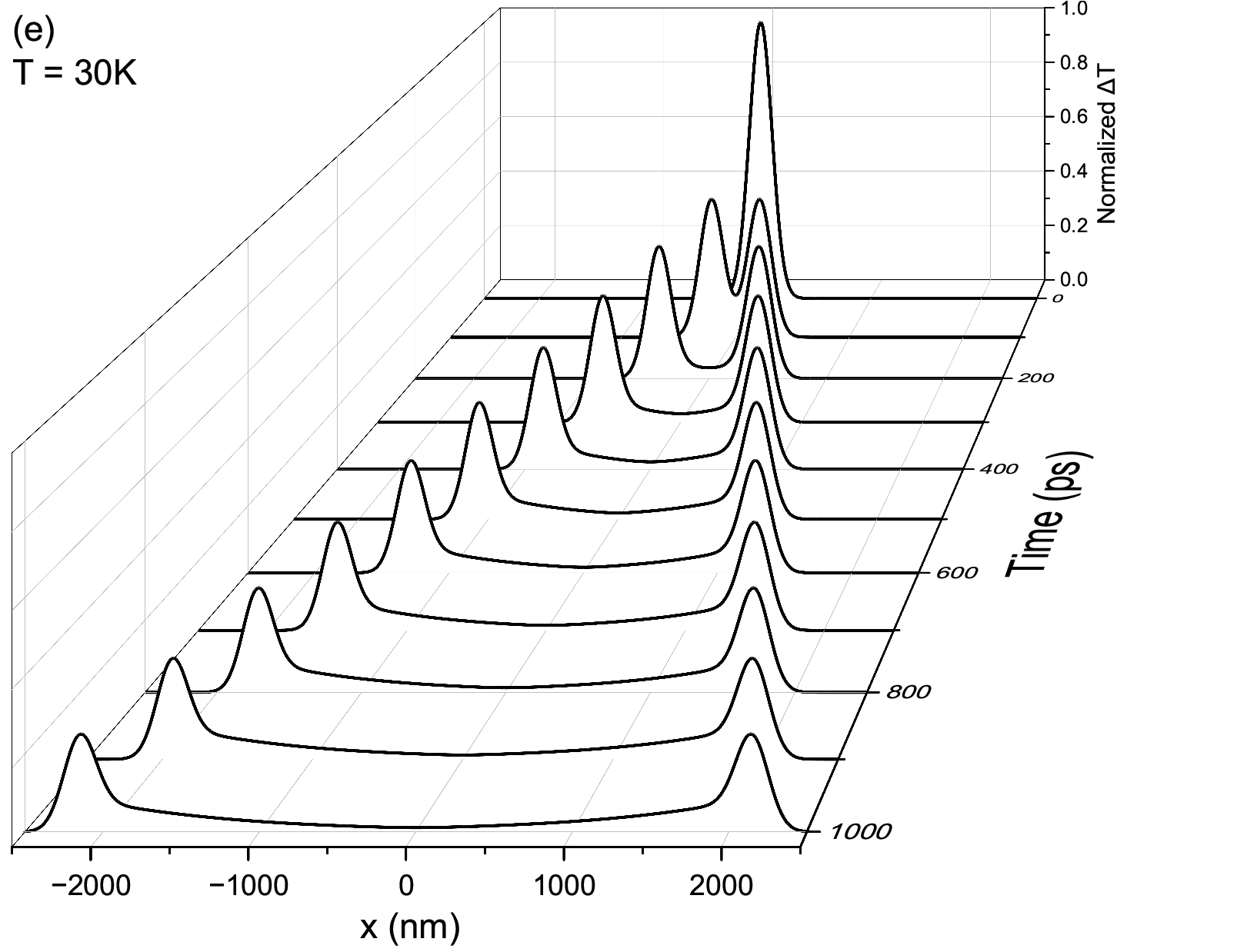}
    \end{minipage}
    \begin{minipage}{0.3\linewidth} 
        \includegraphics[width=\textwidth]{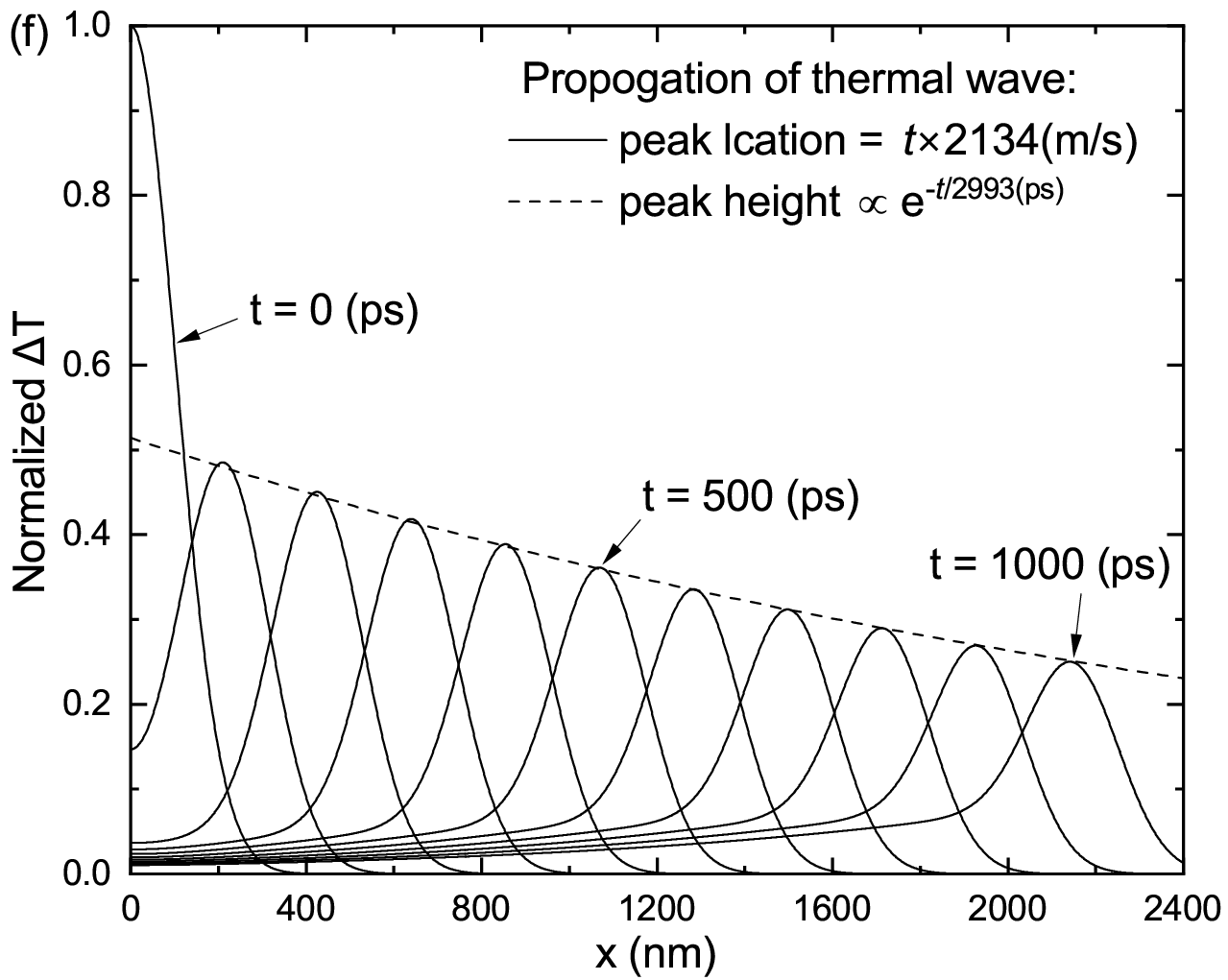}
    \end{minipage}
    \caption{3D plots of the temperature field's evolution over time and space in natural Si crystals at (a) 300K, (b) 200K, (c) 100K, (d) 50K and (e) 30K. The evolution of $T(x,t)$ is solved using Eq. \ref{eq:zHCE_delta}, with the $Z(t)$ from {\bf Fig. \ref{fig:DFT_calculation}} as input.  (f) Snapshots of the temperature profile (solid lines) in natural silicon crystals at 30K, taken every 100ps over a total period of 1000ps. The peak positions of the temperature profiles are linearly fitted as a function of time, yielding a wave propagation speed of 2134 m/s. The decay in peak heights is fitted with an exponential function, with a time constant of 2993~ps. }
    \label{fig:temp_profile}
\end{figure*}

The strong temperature dependence of the first-principles-predicted transport functions, $\overleftrightarrow{Z}(t)$, for Silicon ({\bf Fig.\ref{fig:DFT_calculation}}) leads to distinct dynamics of heat conduction across different temperature regimes. {\bf{Fig. ~\ref{fig:temp_profile} (a–e)}} present the one-dimensional temperature profiles predicted by the Zwanzig theory at various temperatures, ranging from room to cryogenic conditions. The initial temperature profile consists of a uniform background at 300~K, 200~K, 100~K, 50~K, and 30~K, respectively, with an additional spatially localized Gaussian-shaped temperature increase. This localized hotspot has a standard deviation ($\sigma$) of 102.4~nm and is confined within a full width of 1024~nm. 

With a broader initial temperature profile than that shown in Fig.\ref{fig:Z_vs_F}, the non-Gaussian characteristics of the temperature profiles ({\bf Fig.\ref{fig:temp_profile}(a)}) become less pronounced, even during the early stages of heat conduction. The differences observed between {\bf Fig.\ref{fig:Z_vs_F}} and {\bf Fig.\ref{fig:temp_profile}(a)} underscore the strong dependence of transient heat conduction dynamics on the spatial length scale of the initial temperature profiles. Notably, the Zwanzig theory predicts heat conduction behavior that closely resembles Fourier-like diffusion at 300~K when the length scale of temperature profiles exceeds 1~$\mu$m.

As the temperature decreases to 200 K, the transport function $Z(t)$ exhibits a slower decay rate, resulting in more pronounced non-Gaussian features in the predicted temperature profiles, even at timescales extending to 1$\mu$s, as illustrated in {\bf Fig.\ref{fig:temp_profile}(b)}. This slower decay of $Z(t)$ reflects the reduced efficiency of phonon scattering processes at lower temperatures, which amplifies the memory effects captured by the Zwanzig theory. Despite these enhanced memory effects, {\bf Fig.\ref{fig:temp_profile}(b)} shows that diffusion-like behavior characteristic of conventional Fourier heat conduction still dominates the overall dynamics at 200~K in Silicon, particularly over longer timescales.
 
As the temperature further decreases to 100 K ({\bf Fig. \ref{fig:temp_profile}(c)}), wave-like dynamics become increasingly prominent, with deviations from Gaussian diffusion profiles becoming much more pronounced. Initially, these waves retain shapes and widths similar to the initial profile as they propagate in opposite directions. Meanwhile, the heights of these waves decay rapidly, accompanied by the emergence of a broader, Gaussian-like background temperature profile. This broad Gaussian-like component corresponds to conventional diffusion, with its width increasing over time. The wave peaks merge into the broader background profile after 1000~ps.  The difference between the broad profile predicted by our Zwanzig theory and the Gaussian-shaped profile predicted by the conventional Fourier theory remains noticeable  at the sub-micron-second time-scale. In Silicon crystals, the 100~K likely marks as the transition regime with strong interplay between conventional diffusion dynamics and the emerging wave-dynamics. 

As illustrated in {\bf Fig.\ref{fig:temp_profile}(d)}, the Zwanzig theory predicts that wave dynamics dominate heat conduction on nanosecond timescales at 50 K, signaling a breakdown of Fourier theory under such conditions. At 30 K, the steady-state thermal conductivity $\kappa$ of pristine Silicon, free from defects and grain boundaries, exceeds 4000 W/m$\cdot$K. This exceptionally high $\kappa$ arises from the slow rates of collective phonon fluctuation and dissipation processes, as reflected by the eigenmodes of the phonon scattering matrix with large $\gamma^{-1}$ values ({\bf Fig.\ref{fig:DFT_calculation}(b)}). Under these conditions, the Zwanzig theory predicts a pronounced wave propagation pattern for heat conduction ({\bf Fig.\ref{fig:temp_profile}(e)}), closely resembling second sound behavior and marking a complete breakdown of the conventional concept of steady-state thermal conductivity.

To evaluate the propagation speed of second sound-type temperature waves, the time evolution of temperature profiles at 30 K for natural silicon crystals is plotted in {\bf Fig.\ref{fig:temp_profile}(f)}. Our first-principles simulations predict weakly damped second sound wave propagation in silicon at 30 K, with an estimated second sound (ss) speed,  $v_{ss}$, of 2133 m/s and a damping time constant,  $\tau_{ss}$, of 1503~ps. The propagating wave packet experiences minimal damping and is accompanied by a small but detectable diffusive background, indicating the coexistence of wave-like and diffusion-like heat transport. In this ultra-high conductivity regime, heat dissipation is dominated by collective phonon mechanisms, where the relevance of individual phonon lifetimes and mean free paths diminishes. Traditional quasi-ballistic models, which depend on these parameters, fail to capture the observed wave dynamics. By contrast, the Zwanzig theory provides a robust framework for predicting the transition from diffusion-dominated to wave-dominated heat conduction. This capability highlights its utility in advancing our understanding of heat transfer processes in materials with exceptionally high thermal conductivity.

\subsection{TTG Response Functions}
Our discovery of universal onset of wave dynamics in heat conduction, observed across a wide range of temperatures, has important implications for interpreting transient thermal measurements, such as transient thermal grating (TTG) experiments, where the spacing of the grates, $d$, determines the temperature evolution at wave number $q_{0} = 2 \pi/{d}$. The TTG response function predicted by the Fourier theory is given by $\Delta T_{q_{0}}(t) = e^{-q_{0}^{2}D_{0} t }$, where $D_{0}$ represents the steady-state thermal diffusivity. The difference of $\Delta T_{q_{0}}(t)$ between the measured TTG in real materials and the Fourier theory have been previously interpreted as non-Fourier effects in heat conduction\cite{johnson2013direct,ding2022observation}. While standard TTG setups typically use grates with sizes between 0.5 and 100 $\mu m$, recent advancements in TTG spectroscopy with extreme ultraviolet (EUV) and x-ray wavelengths have enabled promising opportunities to probe of heat transfer at the nanometer scale \cite{2019_EUV_TTG}.

\begin{figure*}[h]
\centering
\includegraphics[width=\textwidth]{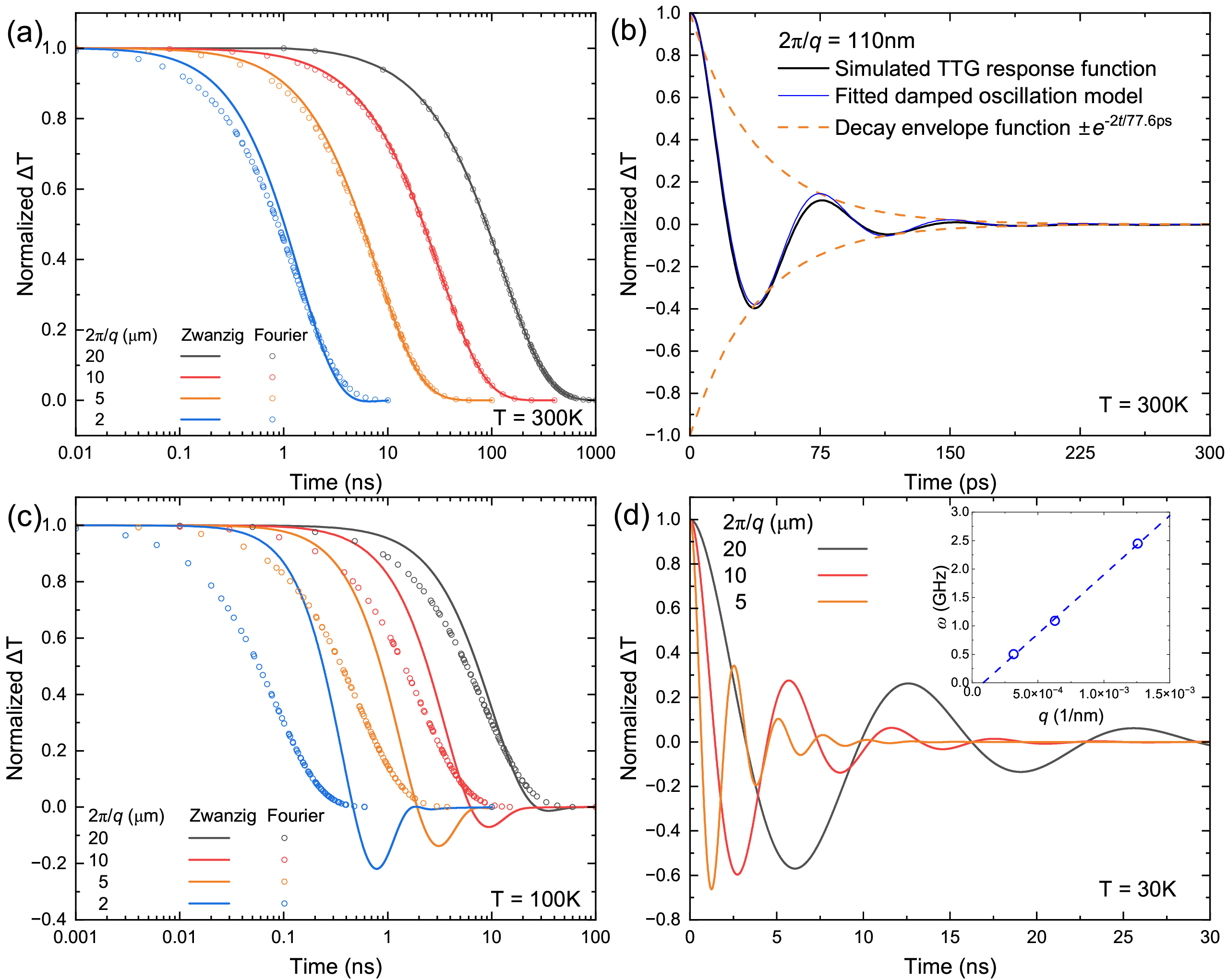}
\caption{Simulated TTG response functions using the newly proposed Zwanzig theory (solid lines) and the conventional Fourier theory (discrete symbols) at 300 K ((a) and (b), 100 K (c), and 30 K (d). Three $\mu$m grates are chosen for 300 K (a), 100 K (c), and 30 K (d), with an additional EUV grate $d$ of 110 nm at 300 K (b). The second sound type of oscillating TTG response function are fitted with a damped oscillation model (DOM): $ e^{-\alpha  t} (cos(\omega_1t)+\alpha/\omega_1 sin(\omega_1t))$, and the harmonic frequency defined as $\omega_0 = (\omega{_1}{^2} + \alpha^2)^{1/2}$. The fitted DOM function at 300 K are shown with the thin blue lines in (b), and the decaying magnitude function are plotted with the red dashed lines. The inset of (d) shows the extrapolated dispersion plot of phonon second sound at 30 K.} \label{fig:ttg}
\end{figure*}

The real components of the $q_{0}$-dependent $\Delta T_{q_{0}}(t)$ functions, as predicted by Zwanzig theory (Eq. \ref{eq:TTG_delta}  and first-principles calculations ({\bf Fig. \ref{fig:DFT_calculation}}), are shown as solid lines in {\bf Fig. \ref{fig:ttg}} for 300 K, 100 K, and 30 K. As expected, Zwanzig theory validates Fourier theory at large length scales, i.e. at the $q_{0} \rightarrow 0$ limit. Meanwhile, in natural silicon crystals and at 300 K, noticeable differences between the two models appear at a small grating spacing $d$ of 2 $\mu$m ({\bf Fig. \ref{fig:ttg}(a)}). More importantly, with an EUV-scale grating $d$ of 110 nm, our simulations confirm the emergence of a distinct oscillating TTG response function resembling phonon second sound (thick black solid line in {\bf Fig. \ref{fig:ttg}(b)}). This numerically simulated EUV-TTG response function fits well with a damped oscillation model, characterized by a magnitude function $e^{- t/{38.8 \text{ps}}}$ and an oscillating function with a harmonic angular frequency $\omega_{0}$ of 87.67 GHz (thin blue solid line in {\bf Fig. \ref{fig:ttg}(b)}). Recent EUV-TTG experiments on a silicon crystalline membrane at 300 K, using a 110 nm grating \cite{2019_EUV_TTG}, revealed an unexpected complex TTG response. The data show a fast oscillatory component, with frequencies of several hundred GHz or higher, superimposed on a much slower, exponentially decaying function. Notably, unlike in the small $q_0$ limit, the measured exponential decay deviates significantly from the Fourier theory, $e^{-q_{0}^{2}D_{0}t}$. Instead, the measured decay function with decay constant at 36.5 ps closely resembles our simulated magnitude function of the EUV-TTG response function with decay constant at 38.8 ps. The origin of the experimentally observed oscillation feature, however, remains unsolved at this stage.

Although detecting deviation from Fourier theory and the emergence of second sound type TTG response functions in Si crystals at 300 K requires advanced experimental setups in the EUV regime, non-Fourier signatures in heat conduction are easier to detect with standard micron meter TTG setups as the temperature decreases. {\bf Fig. \ref{fig:ttg}(c)} shows that Fourier theory becomes significantly inaccurate at 100 K, even with a grating 20 $\mu $m. At 30 K ({\bf Fig. \ref{fig:ttg}(d)}), Fourier theory completely breaks down, and phonon second sound oscillations are observed for gratings of $d = 20\, \mu$m and shorter. The dispersion relation of the phonon second sound waves at 30 K is shown in the inset of {\bf Fig. \ref{fig:ttg}(d)}, with an extracted speed around 2100 m/s. Notably, most reported TTG experiments have been conducted on thin membrane samples, and for silicon, TTG responses have been specifically reported for nano-porous Si membranes\cite{vega2016thermal}. Our simulated TTG response functions, based on the $Z(t)$ function of bulk crystalline Si, are intended to illustrate temperature and $q_{0}$-dependence rather than provide a quantitative comparison with experimentally measured TTG responses from nano-porous Si membranes.
\section{Summary and Outlook}
This work introduces a time-domain Zwanzig theory for transient heat conduction, offering a unified framework that extends Fourier’s law and deepens our understanding of heat transport in homogeneous materials across a broad range of temporal and spatial scales. Central to this framework is the time-domain transport function, $\overleftrightarrow{Z}(t)$, rigorously defined in terms of equilibrium time-correlation functions of heat fluxes. This $\overleftrightarrow{Z}(t)$-centered theory generalizes the classical concept of thermal conductivity beyond the quasi-steady regime, capturing memory effects and wave-diffusion crossover dynamics without reliance on mesoscopic constructs, such as phonon hydrodynamic drift velocity. By bridging the gap between steady-state transport coefficients and time-dependent heat flux, the theory establishes a set of governing equations at the continuum level, providing a robust framework for modeling thermal transport at micrometer to nanometer length scales and microsecond to picosecond timescales, where conventional Fourier diffusion theory fails.

Building on Zwanzig’s statistical theory of irreversible processes, this study advances the formalism by integrating it into a continuum-level governing equation, thereby extending its applicability to ultrafast and nanoscale heat conduction, which is a domain that has remained largely unexplored in prior discussions of Zwanzig’s fundamental theory. Our derivation of a phonon gas' time-domain transport function, $\overleftrightarrow{Z}(t)$, from the time-dependent phonon Boltzmann transport equation provides an intuitive application of the general Zwanzig theory. This establishes a unified framework that connects and generalizes existing models addressing memory effects in phonon heat fluxes, including the hydrodynamic Guyer-Krumhansl equation or bulk phonon gases and the empirical MVC model. Notably, our first-principles calculations of $\overleftrightarrow{Z}(t)$ for natural silicon crystals validate the predictive power and accuracy of this approach, underscoring its potential to advance our understanding of transient heat conduction.

The time-domain formalism introduces significant computational advantages through a mode-specific algorithm that approximates $\overleftrightarrow{Z}(t)$ as a finite sum of exponential functions. This approximation enables efficient continuum-level simulations of temporally pulse-like thermal signals, achieving high numerical accuracy while substantially reducing computational complexity.  Notably, the widely adopted heat telegrapher equations emerge as limiting cases within this framework, highlighting its versatility and capacity to unify diverse approaches to modeling heat conduction.

Our time-domain simulations demonstrate that heat conduction universally initiates with wave-like propagation at finite speeds, transitioning to diffusion-dominated behavior over time. This crossover is governed by the spatial confinement of the initial temperature profile and the material-specific time-domain transport function, $\overleftrightarrow{Z}(t)$. Localized heating patterns or temperature profiles can be decomposed into spatially periodic components characterized by wave numbers $\vec{q}$. Components with small wave numbers $\vec{q}$ evolve diffusively, while those with larger $\vec{q}$ propagate as waves, highlighting second-sound waves as a universal mechanism of heat conduction across materials. This wave-diffusion crossover theory provides a unified and fundamental interpretation of length-scale-dependent thermal transport phenomena, bridging the gap between traditional diffusion models and wave-dominated dynamics.

Numerical simulations of TTG response functions in bulk silicon validate the proposed framework, effectively capturing the interplay between wave-like and diffusive heat transport dynamics. The theory further suggests that advancements in TTG techniques could enable direct measurement or calibration of the time-domain transport function, $\overleftrightarrow{Z}(t)$, bridging the gap between theoretical predictions and experimental methodologies. Additionally, the Zwanzig theory provides a robust foundation for refining models of other transient techniques, such as time-domain thermal reflectance (TDTR), in the non-Fourier regime. While current applications are limited by incomplete knowledge of $Z(t)$ in metals and metal/thin-film interfaces, the theory predicts a modulation-frequency-dependent thermal diffusivity, $D(\omega)$, which could replace the steady-state value, $D(\omega = 0)$, in TDTR models\cite{cahill2004analysis}. Optimistically,, the time-domain formalism offers a direct pathway for experimental validation through transient thermal techniques such as TTG spectroscopy and TDTR measurements, paving the way for a deeper understanding of heat conduction in complex materials.

Finally, the time-domain Zwanzig framework transcends heat conduction, offering applicability to other transient transport phenomena, including mass, charge, and spin conduction in condensed matter systems. By incorporating time-domain transport functions and the mode-specific algorithm, it provides a versatile and unified approach to understanding transient dynamics in complex materials, paving the way for advances in fundamental research and technological applications.

\section{Method}\label{sec:method}
\subsection{First-principle calculations}
We computed the steady-state lattice thermal conductivity ($\kappa$) and the equilibrium time-correlation functions of phonon heat flux for bulk silicon (Si) crystals using first-principles lattice dynamics. Harmonic and third-order anharmonic force constants were derived via density functional theory (DFT) calculations performed with the Vienna Ab initio Simulation Package (VASP) \cite{Kresse1996}. The Perdew-Burke-Ernzerhof (PBE) exchange-correlation functional within the generalized gradient approximation (GGA) was employed. A plane-wave cutoff energy of 245.345 eV was used, and Brillouin zone integration was carried out using a $31 \times 31 \times 31$ $\vec{k}$-point grid for a two-atom face-centered cubic unit cell, ensuring convergence of the unit cell energy and vibrational properties. Harmonic phonon spectra were calculated using Phonopy \cite{phonopy}, while linearized phonon scattering matrices were generated with Phono3py \cite{phono3py}. For the solution of the phonon Boltzmann transport equation, a $21 \times 21 \times 21$ $\vec{q}$-point grid was utilized, encompassing 55,566 phonon modes.

The force constants were extracted using a 216-atom cubic supercell model of silicon, with a lattice constant of 5.473 \text{\AA}. This supercell-based approach in VASP has been extensively validated for its robustness in simulating phonon scattering processes across a variety of dielectric crystals with diverse structural and compositional complexities, including MgO and $\text{MgSiO}_3$ under high-pressure conditions \cite{tang2009pressure,Tang4539,tang2014thermal}. Both three-phonon scattering processes and phonon scattering due to isotope mass disorder were explicitly accounted for in this study to ensure accurate modeling of thermal transport properties.

\subsection{Numerical simulation of transient heat conduction}
The time evolution of a Gaussian heat pulse in silicon was simulated by numerically solving Eqs.(\ref{eq:zHCE_delta}) using a finite difference method. The computational domain had a total length of $L = 5120 \text{nm}$, discretized into 5120 mesh points with a spatial resolution of $\Delta x = 1~\text{nm}$. Temporal resolution was achieved using a time step of $\Delta t = 1~\text{fs}$, with a total of $10^6$ time steps covering a simulation time of 1000~ps. The width of the initial heat pulse was adjusted based on detailed case-specific simulations.

For the simulation of TTG response functions at varying temperatures (\ref{eq:TTG_delta}, the total simulation time was modified to correspond to the grating period ($d$) under consideration. The specific values for $d$ and associated simulation parameters are provided in Table~\ref{table:TTG_d}.

\begin{table}[h]
\caption{\label{table:TTG_d}
Total simulation time for various grating period, $d$, at different temperatures}
\begin{ruledtabular}
\begin{tabular}{ccccc}
$T$(K)  &$d$ ($\mu \text{m}$) & $t_{\text{total}} $(ns) \\
\hline
30& 5 & 50  \\
& 10 & 100 \\
& 20 & 100 \\
\hline
100& 2 & 10  \\
& 5 & 10 \\
& 10 & 100 \\
& 20 & 100 \\
\hline
300& 0.11 & 0.5  \\
& 2 & 10 \\
& 5 & 100 \\
& 10 & 400 \\
& 20 & 1000 \\
\end{tabular}
\end{ruledtabular}
\end{table}


\begin{acknowledgments}
This work was authored in part by Auburn University, and the National Renewable Energy Laboratory, operated by Alliance for Sustainable Energy, LLC, for the U.S. Department of Energy (DOE) under Contract No. DE-AC36-08GO28308. Funding provided by Auburn University and by the U.S. Department of Energy Office of Science Energy Earthshots Initiative as part of the of Degradation Reactions in Electrothermal Energy Storage (DEGREES) project. The views expressed in the article do not necessarily represent the views of the DOE or the U.S. Government. The U.S. Government retains and the publisher, by accepting the article for publication, acknowledges that the U.S. Government retains a nonexclusive, paid-up, irrevocable, worldwide license to publish or reproduce the published form of this work, or allow others to do so, for U.S. Government purposes. J.D. and Y.Z. also acknowledge discussions with Wencan Jin (Auburn), David A. Drabold (Ohio), Shannon K. Yee (Georgia Tech), Shuang Cui (UT Dallas). The numerical calculations were carried out using the Auburn University Easley high-performance computer. 
\end{acknowledgments}

\appendix
\section{\label{sec:linear}Linearized time-dependent phonon Boltzmann transport equation}
Our microscopic statistical theory pf transient heat current in a phonon gas is established using the time-dependent phonon Boltzmann transport equation (td-phBTE):
\begin{equation}
\label{eq_phBTE_1}
 \frac{dn(\alpha)}{dt} =  \frac{\partial n(\vec{r},\alpha,t)}{\partial t} + \vec{\nabla}_{r} n(\vec{r},\alpha, t) \cdot \vec{v}_{\alpha} = \left.\frac{dn(\vec{r}, \alpha,t)}{dt} \right|_{\text{relax}},  
\end{equation}
where $\alpha$ denotes both the wavevector $\vec{q}$ in reciprocal space and the phonon branch index $b$, i.e., $\alpha = (\vec{q}, b)$,  and $n(\vec{r}, \alpha, t)$ is the non-equilibrium phonon occupation number for the mode $\alpha$  at position $\vec{r}$ and time $t$. We consider a perfect periodic crystal subjected to an externally imposed, temporally varying, and spatially uniform temperature gradient, $\vec{\nabla}_r T(t)$, where the phonon distribution gradient can be approximated as:
\begin{equation}
\label{eq:gradient}
\vec{\nabla}_r n(\vec{r}, \alpha,  t) \approx \vec{\nabla}_r T(t) \cdot \frac{\partial n_{0}(\alpha)}{\partial T},
\end{equation}
with $n_{0}(\alpha) = 1/\big({e^{\frac{\hbar \omega_{\alpha}}{k_B T}} - 1}\big)$ being the equilibrium phonon occupation number at temperature $T$ for the $\alpha$-th mode with phonon frequency $\omega_{\alpha}$. The temperature derivative of the equilibrium phonon occupation number is given by:
\begin{equation}
\frac{\partial n_{0}(\alpha)}{\partial T} = \frac{\hbar \omega_{\alpha}}{4 k_B T^2 \sinh^2\left(\frac{\hbar \omega_{\alpha}}{2 k_B T}\right)}.
\end{equation}

Under the condition of spatially uniform temperature gradient $\vec{\nabla}_r T(t)$, the td-phBTE for the non-equilibrium phonon occupation numbers are no longer functions of $\vec{r}$: 
\begin{equation}
\label{eq_phBTE_2}
 \frac{dn(\alpha)}{dt} =  \frac{\partial n(\alpha,t)}{\partial t} +   \frac{\hbar \omega_{\alpha}}{4 k_B T^2 \sinh^2\left(\frac{\hbar \omega_{\alpha}}{2 k_B T}\right)}  \vec{\nabla}_{r} T(t) \cdot \vec{v}_{\alpha} = \left.\frac{dn(\alpha,t)}{dt} \right|_{\text{relax}},  
\end{equation}

We further adopt a linearized approximation for phonon relaxation dynamics and replace the integration over reciprocal $\vec{q}$-space by a summation over discrete $\vec{q}$-points, , where the rates of phonon scattering are expressed in terms of the symmetric and semi-positive-definite linearized phonon scattering matrix, $L$ \cite{zeng_and_dong_2019_vibFPE}:
\begin{equation}
\label{eq_lpsm}
\left. \frac{d n(\alpha,t)}{d t} \right|_{\text{relax}} \approx - \sum_{\beta} L_{\alpha \beta} \cdot \frac{\sinh \left(\frac{\hbar \omega_{\beta}}{2 k_B T_0} \right)}{\sinh \left(\frac{\hbar \omega_{\alpha}}{2 k_B T_0} \right)} \cdot (n(\beta, t) - n_{0}(\beta)),
\end{equation}
where $n_{\alpha}(t)$ and $n_{\beta}(t)$ are the time-dependent phonon occupation numbers for modes $\alpha$ and $\beta$.

Using this approximation, we can rewrite the linearized td-phBTE as:
\begin{equation}
\label{eq_td-phBTE}
\begin{aligned}
\frac{\partial n(\alpha,t)}{\partial t} + \frac{\hbar \omega_{\alpha}}{4 k_B T_0^2} \frac{\vec{\nabla}_r T(t) \cdot \vec{v}_{\alpha}}{\sinh^2 \left(\frac{\hbar \omega_{\alpha}}{2 k_B T_0}\right)} = 
- \sum_{\beta} L_{\alpha \beta} \cdot \frac{\sinh \left(\frac{\hbar \omega_{\beta}}{2 k_B T_0}\right)}{\sinh \left(\frac{\hbar \omega_{\alpha}}{2 k_B T_0}\right)} \cdot \big(n(\beta,t) - n_{0}(\beta)\big).
\end{aligned}
\end{equation}

\section{\label{sec:solution} Analytical solution of td-phBTE}
We now introduce a new set of variables, $\psi_{\alpha}(t) \equiv 2 \sinh \left(\frac{\hbar \omega_{\alpha}}{2 k_B T_0}\right) \cdot \big(n(\alpha,t) - n_{0}(\alpha)\big)$, and re-formulate the linearized td-phBTE as:
\begin{equation}
\label{eq-ldphbte_2}
\frac{d}{dt} \psi_{\alpha} (t) + \frac{\hbar \omega_{\alpha} \Vec{v}_{\alpha}  }{2k_B T_{0}^2 \sinh \left(\frac{\hbar \omega_{\alpha}}{2 k_{B} T_{0}} \right)} \Vec{\nabla}_r T(t) = - \sum_{\beta} L_{\alpha \beta} \psi_{\beta} (t).    
\end{equation}
We diagonalize the semi-positively defined, real, symmetric $N_{\text{mode}} \times N_{\text{mode}}$ matrix $L$ to obtain a set of $N_{\text{mode}}$ orthonormal eigenvectors $\Vec{u}_{\lambda}$ and their corresponding eigenvalues $\gamma_{\lambda}$ \cite{zeng_and_dong_2019_vibFPE}. These satisfy the eigenvalue equation $\sum_{\beta=1}^{N_{\text{mode}}} L_{\alpha \beta} u_{\lambda \beta} = \gamma_{\lambda} u_{\lambda \alpha}$, where $\lambda$ indexes the eigenvectors. To simplify further, we introduce another set of variables, $\eta_{\lambda}(t) \equiv \sum_{\alpha=1}^{N_{\text{mode}}} u_{\lambda \alpha} \psi_{\alpha}(\Vec{r}, t)$, allowing us to rewrite the linearized td-phBTE in terms of the linearized phonon scattering matrix’s eigenvectors as a new orthonormal basis set:
\begin{equation}
\label{eq-ldphbte_3}
\frac{d}{dt} \eta_{\lambda} (t)   + \gamma_{\lambda} \eta_{\lambda} (t) = -  \sum_{\alpha} \frac{u_{\lambda \alpha} \hbar \omega_{\alpha} \Vec{v}_{\alpha} \cdot}{2k_B T_{0}^2 \sinh \left(\frac{\hbar \omega_{\alpha}}{2 k_{B} T_{0}} \right)} \Vec{\nabla}_r T(t), 
\end{equation}
or
\begin{equation}
\label{eq-ldphbte_4}
\frac{d}{dt} \big( e^{\gamma_{\lambda} t} \cdot \eta_{\lambda} (t)  \big) = -  \sum_{\alpha} \frac{u_{\lambda \alpha} \hbar \omega_{\alpha} \Vec{v}_{\alpha} \cdot}{2k_B T_{0}^2 \sinh \left(\frac{\hbar \omega_{\alpha}}{2 k_{B} T_{0}} \right)} e^{\gamma_{\lambda} t} \cdot \Vec{\nabla}_r T(t), 
\end{equation}
The first-order differential equation above can be solved analytically as follows:
\begin{equation}
    \label{eq_solution}
    \eta_{\lambda}  (t)  =   - \sum_{\alpha} \frac{u_{\lambda \alpha} \hbar \omega_{\alpha} \Vec{v}_{\alpha} \cdot}{2k_B T_{0}^2 \sinh \left(\frac{\hbar \omega_{\alpha}}{2 k_{B} T_{0}} \right)}  \int_{-\infty}^{t}dt' e^{-\gamma_{\lambda} (t-t')} \Vec{\nabla}_r T(t') .
\end{equation}
The time-evolution of the non-equilibrium phonon occupation number for the $\alpha$-th phonon mode is derived as:
\begin{equation}
    \label{eq_n_alpha_of_t}
    \begin{split}
   & n(\alpha,t) \\
   = &n_{0}(\alpha) + \sum_{\lambda = 1}^{N_{mode}} \frac{u_{\lambda \alpha} \eta_{\lambda}  (t) }{2 \sinh \left(\frac{\hbar \omega_{\alpha}}{2 k_B T_0}\right) }  \\
    = & n_{\alpha}^{0}(t)  - \sum_{\lambda, \beta = 1}^{N_{mode}} \frac{u_{\lambda \alpha} u_{\lambda \beta} \hbar \omega_{\beta} \Vec{v}_{\beta} \cdot}{4k_B T_{0}^2 \sinh \left(\frac{\hbar \omega_{\alpha}}{2 k_{B} T_{0}} \right)  \sinh \left(\frac{\hbar \omega_{\beta}}{2 k_{B} T_{0}} \right) }  \int_{-\infty}^{t}dt' e^{-\gamma_{\lambda} (t-t')} \Vec{\nabla}_r T(t') .
    \end{split}
\end{equation}

\section{\label{sec:gHTE}Generalized heat telegrapher equation}
While heat diffusion processes are traditionally modeled using the parabolic heat diffusion equation (Eq.~\ref{eq:diffusion}), wave-like heat transfer, often referred to as second sound waves, requires a fundamentally different mathematical framework. A generalized approach that can describe heat conduction transitioning between diffusive and wave-like behaviors is embodied in the Heat Telegrapher Equation (HTE):
\begin{equation}
\label{eq:HTE}
    \frac{\partial^2 }{\partial t^2}T(\vec{r}, t) + \frac{1}{\tau_{ss}}  \frac{\partial }{\partial t}T(\vec{r}, t) -  v^{2}_{ss} \nabla^{2}_{r} T(\vec{r}, t)= 0, 
\end{equation}
where  $v_{ss}$ denotes the speed of sound sound waves, and  $\tau_{ss}$  represents a characteristic damping time.

The HTE is a second-order partial differential equation that incorporates features of both the parabolic heat diffusion equation and the hyperbolic wave equation. The second time derivative describes wave-like propagation with a finite speed, while the first time derivative introduces damping effects that result in an exponential decay of the temperature fields. In the high-damping or low-frequency limit ($\tau_{ss} \to 0$), the HTE simplifies to a form analogous to the diffusion equation, which characterizes the slow, diffusive spreading of temperature fields. Conversely, in the low-damping or high-frequency regime ($\tau_{ss} \to \infty$), the HTE approaches the wave equation, capturing under-damped wave propagation with distinct wave fronts. This dual nature allows the HTE to serve as a unifying framework that interpolates between purely diffusive and purely wave-like heat conduction dynamics, making it a powerful tool for understanding heat transfer across a wide range of temporal and spatial scales.

We now show how the Zwanzig heat conduction equation (Eq.\ref{eq:zHCE}) can be reformulated into a generalized heat telegrapher equation (gHTE). To begin, we take the time derivative of both sides of the first equation in Eq.\ref{eq:zHCE_mode}:
\begin{equation}
\label{eq:t-deriv_1}
\frac{\partial^2}{\partial t^2}T(\vec{r},t) = -\frac{1}{\rho C} \vec{\nabla}_{r} \cdot \big( \sum_{\lambda =1}^{N_{\text{mode}}}  \frac{\partial}{\partial t} \vec{j}_{\lambda}(\vec{r},t)\big). 
\end{equation}

Substituting the time derivative of $\vec{j}{\lambda}(\vec{r},t)$ from the second equation in Eq.~\ref{eq:zHCE_mode}, we obtain: 
\begin{equation}
\label{eq:t-deriv_2}
\frac{\partial^2}{\partial t^2}T(\vec{r},t) =  \frac{1}{\rho C} \vec{\nabla}_{r} \cdot \big( \sum_{\lambda =1}^{N_{\text{mode}}} \overleftrightarrow{\zeta}_{\lambda 0}  \big) \cdot  \vec{\nabla}_{r}  T(\vec{r},t)  + \frac{1}{\rho C} \sum_{\lambda =1}^{N_{\text{mode}}} \big( \gamma_{\lambda} \vec{\nabla}_{r} \cdot \vec{j}_{\lambda}(\vec{r},t)\big). 
\end{equation}

This can be reformulated as:
\begin{equation}
\label{eq:t-deriv_3}
\frac{\partial^2}{\partial t^2}T(\vec{r},t) =  \frac{1}{\rho C} \vec{\nabla}_{r} \cdot \big( \sum_{\lambda =1}^{N_{\text{mode}}} \overleftrightarrow{\zeta}_{\lambda 0}  \big) \cdot  \vec{\nabla}_{r}  T(\vec{r},t)  -  \Gamma(\vec{r},t) \frac{\partial}{\partial t}T(\vec{r},t),  
\end{equation}
where $\Gamma(\vec{r}, t)$ is a a mode-flux-weighted damping term, defined as:  
\begin{equation}
\label{eq:Gamma}
\begin{split}
\Gamma (\vec{r}, t)   \equiv & \frac{\sum_{\lambda=1}^{M}\gamma_{\lambda} \vec{\nabla}_{r}  \vec{j}_{\lambda}(\vec{r}, t)}{\vec{\nabla}_{r}  \vec{j}(\vec{r}, t) } \\ 
= - &  \frac{\sum_{\lambda=1}^{M}\gamma_{\lambda} \vec{\nabla}_{r}  \vec{j}_{\lambda}(\vec{r}, t)}{\rho C \frac{\partial}{\partial t}T(\vec{r},t)} 
\end{split}
\end{equation}
Here,  $\Gamma(\vec{r}, t)$ is generally not a constant but a function of both space and time. Rearranging Eq. \ref{eq:t-deriv_3} and setting $ \overleftrightarrow{Z}(0)=\sum_{\lambda =1}^{N_{\text{mode}}} \overleftrightarrow{\zeta}_{\lambda 0} $, we arrive at the generalized HTE:
\begin{equation}
\label{eq:gHTE}
    \frac{\partial^2 }{\partial t^2}T(\vec{r}, t) + \Gamma(\vec{r}, t)  \frac{\partial }{\partial t}T(\vec{r}, t) -  \vec{\nabla}_{r} \cdot \big(\frac{\overleftrightarrow{Z}(0)}{\rho C} \cdot \vec{\nabla}_{r} T(\vec{r}, t) \big)= 0. 
\end{equation}

For the limiting case of $\text{M}=1$, corresponding to the empirical MCV model or the hydrodynamic GKE for a bulk crystal, $\Gamma(\vec{r},t)$ reduces to a constant, $\Gamma(\vec{r},t) = 1/\tau_{ss}$. For isotropic materials, where $\frac{\overleftrightarrow{Z}(0)}{\rho C} = v^{2}_{ss} \overleftrightarrow{I}$, the gHTE simplifies to the conventional HTE shown in Eq.~\ref{eq:HTE}, with $v_{ss} = \sqrt{\frac{D}{\tau_{ss}}}$.

\bibliography{main}

\begin{thebibliography}{85}%
\makeatletter
\providecommand \@ifxundefined [1]{%
 \@ifx{#1\undefined}
}%
\providecommand \@ifnum [1]{%
 \ifnum #1\expandafter \@firstoftwo
 \else \expandafter \@secondoftwo
 \fi
}%
\providecommand \@ifx [1]{%
 \ifx #1\expandafter \@firstoftwo
 \else \expandafter \@secondoftwo
 \fi
}%
\providecommand \natexlab [1]{#1}%
\providecommand \enquote  [1]{``#1''}%
\providecommand \bibnamefont  [1]{#1}%
\providecommand \bibfnamefont [1]{#1}%
\providecommand \citenamefont [1]{#1}%
\providecommand \href@noop [0]{\@secondoftwo}%
\providecommand \href [0]{\begingroup \@sanitize@url \@href}%
\providecommand \@href[1]{\@@startlink{#1}\@@href}%
\providecommand \@@href[1]{\endgroup#1\@@endlink}%
\providecommand \@sanitize@url [0]{\catcode `\\12\catcode `\$12\catcode `\&12\catcode `\#12\catcode `\^12\catcode `\_12\catcode `\%12\relax}%
\providecommand \@@startlink[1]{}%
\providecommand \@@endlink[0]{}%
\providecommand \url  [0]{\begingroup\@sanitize@url \@url }%
\providecommand \@url [1]{\endgroup\@href {#1}{\urlprefix }}%
\providecommand \urlprefix  [0]{URL }%
\providecommand \Eprint [0]{\href }%
\providecommand \doibase [0]{http://dx.doi.org/}%
\providecommand \selectlanguage [0]{\@gobble}%
\providecommand \bibinfo  [0]{\@secondoftwo}%
\providecommand \bibfield  [0]{\@secondoftwo}%
\providecommand \translation [1]{[#1]}%
\providecommand \BibitemOpen [0]{}%
\providecommand \bibitemStop [0]{}%
\providecommand \bibitemNoStop [0]{.\EOS\space}%
\providecommand \EOS [0]{\spacefactor3000\relax}%
\providecommand \BibitemShut  [1]{\csname bibitem#1\endcsname}%
\let\auto@bib@innerbib\@empty
\bibitem [{\citenamefont {Fourier}(1888)}]{fourier1888theorie}%
  \BibitemOpen
  \bibfield  {author} {\bibinfo {author} {\bibfnamefont {J.~B.~J.}\ \bibnamefont {Fourier}},\ }\href@noop {} {\emph {\bibinfo {title} {Th{\'e}orie analytique de la chaleur}}},\ Vol.~\bibinfo {volume} {1}\ (\bibinfo  {publisher} {Gauthier-Villars},\ \bibinfo {year} {1888})\BibitemShut {NoStop}%
\bibitem [{\citenamefont {Carslaw}\ and\ \citenamefont {Jaeger}(1959)}]{carslaw1959conduction}%
  \BibitemOpen
  \bibfield  {author} {\bibinfo {author} {\bibfnamefont {H.}~\bibnamefont {Carslaw}}\ and\ \bibinfo {author} {\bibfnamefont {J.}~\bibnamefont {Jaeger}},\ }\href@noop {} {\emph {\bibinfo {title} {Conduction of Heat in Solids}}},\ \bibinfo {edition} {2nd}\ ed.\ (\bibinfo  {publisher} {Oxford University Press},\ \bibinfo {address} {Oxford},\ \bibinfo {year} {1959})\BibitemShut {NoStop}%
\bibitem [{\citenamefont {Ziman}(2001)}]{ziman2001electrons}%
  \BibitemOpen
  \bibfield  {author} {\bibinfo {author} {\bibfnamefont {J.~M.}\ \bibnamefont {Ziman}},\ }\href@noop {} {\emph {\bibinfo {title} {Electrons and phonons: the theory of transport phenomena in solids}}}\ (\bibinfo  {publisher} {Oxford University Press},\ \bibinfo {year} {2001})\BibitemShut {NoStop}%
\bibitem [{\citenamefont {Green}(1954)}]{green1954markoff}%
  \BibitemOpen
  \bibfield  {author} {\bibinfo {author} {\bibfnamefont {M.~S.}\ \bibnamefont {Green}},\ }\href@noop {} {\bibfield  {journal} {\bibinfo  {journal} {J. Chem. Phys.}\ }\textbf {\bibinfo {volume} {22}},\ \bibinfo {pages} {398} (\bibinfo {year} {1954})}\BibitemShut {NoStop}%
\bibitem [{\citenamefont {Callaway}(1959)}]{callaway1959model}%
  \BibitemOpen
  \bibfield  {author} {\bibinfo {author} {\bibfnamefont {J.}~\bibnamefont {Callaway}},\ }\href@noop {} {\bibfield  {journal} {\bibinfo  {journal} {Physical Review}\ }\textbf {\bibinfo {volume} {113}},\ \bibinfo {pages} {1046} (\bibinfo {year} {1959})}\BibitemShut {NoStop}%
\bibitem [{\citenamefont {Zwanzig}(1965{\natexlab{a}})}]{zwanzig1965time}%
  \BibitemOpen
  \bibfield  {author} {\bibinfo {author} {\bibfnamefont {R.}~\bibnamefont {Zwanzig}},\ }\href@noop {} {\bibfield  {journal} {\bibinfo  {journal} {Annual Review of Physical Chemistry}\ }\textbf {\bibinfo {volume} {16}},\ \bibinfo {pages} {67} (\bibinfo {year} {1965}{\natexlab{a}})}\BibitemShut {NoStop}%
\bibitem [{\citenamefont {Guyer}\ and\ \citenamefont {Krumhansl}(1966{\natexlab{a}})}]{guyer1966thermal}%
  \BibitemOpen
  \bibfield  {author} {\bibinfo {author} {\bibfnamefont {R.}~\bibnamefont {Guyer}}\ and\ \bibinfo {author} {\bibfnamefont {J.}~\bibnamefont {Krumhansl}},\ }\href@noop {} {\bibfield  {journal} {\bibinfo  {journal} {Physical Review}\ }\textbf {\bibinfo {volume} {148}},\ \bibinfo {pages} {778} (\bibinfo {year} {1966}{\natexlab{a}})}\BibitemShut {NoStop}%
\bibitem [{\citenamefont {Majumdar}\ \emph {et~al.}(1997)\citenamefont {Majumdar}, \citenamefont {Tien},\ and\ \citenamefont {Gemer}}]{majumdar1997microscale}%
  \BibitemOpen
  \bibfield  {author} {\bibinfo {author} {\bibfnamefont {A.}~\bibnamefont {Majumdar}}, \bibinfo {author} {\bibfnamefont {C.}~\bibnamefont {Tien}}, \ and\ \bibinfo {author} {\bibfnamefont {F.}~\bibnamefont {Gemer}},\ }\href@noop {} {\bibfield  {journal} {\bibinfo  {journal} {Microscale Energy Transport}\ ,\ \bibinfo {pages} {3}} (\bibinfo {year} {1997})}\BibitemShut {NoStop}%
\bibitem [{\citenamefont {Chen}(2005)}]{chen2005nanoscale}%
  \BibitemOpen
  \bibfield  {author} {\bibinfo {author} {\bibfnamefont {G.}~\bibnamefont {Chen}},\ }\href@noop {} {\emph {\bibinfo {title} {Nanoscale energy transport and conversion: a parallel treatment of electrons, molecules, phonons, and photons}}}\ (\bibinfo  {publisher} {Oxford university press},\ \bibinfo {year} {2005})\BibitemShut {NoStop}%
\bibitem [{\citenamefont {Dubi}\ and\ \citenamefont {Di~Ventra}(2011)}]{dubi2011colloquium}%
  \BibitemOpen
  \bibfield  {author} {\bibinfo {author} {\bibfnamefont {Y.}~\bibnamefont {Dubi}}\ and\ \bibinfo {author} {\bibfnamefont {M.}~\bibnamefont {Di~Ventra}},\ }\href@noop {} {\bibfield  {journal} {\bibinfo  {journal} {Reviews of Modern Physics}\ }\textbf {\bibinfo {volume} {83}},\ \bibinfo {pages} {131} (\bibinfo {year} {2011})}\BibitemShut {NoStop}%
\bibitem [{\citenamefont {Li}\ \emph {et~al.}(2012)\citenamefont {Li}, \citenamefont {Ren}, \citenamefont {Wang}, \citenamefont {Zhang}, \citenamefont {H{\"a}nggi},\ and\ \citenamefont {Li}}]{li2012colloquium}%
  \BibitemOpen
  \bibfield  {author} {\bibinfo {author} {\bibfnamefont {N.}~\bibnamefont {Li}}, \bibinfo {author} {\bibfnamefont {J.}~\bibnamefont {Ren}}, \bibinfo {author} {\bibfnamefont {L.}~\bibnamefont {Wang}}, \bibinfo {author} {\bibfnamefont {G.}~\bibnamefont {Zhang}}, \bibinfo {author} {\bibfnamefont {P.}~\bibnamefont {H{\"a}nggi}}, \ and\ \bibinfo {author} {\bibfnamefont {B.}~\bibnamefont {Li}},\ }\href@noop {} {\bibfield  {journal} {\bibinfo  {journal} {Reviews of Modern Physics}\ }\textbf {\bibinfo {volume} {84}},\ \bibinfo {pages} {1045} (\bibinfo {year} {2012})}\BibitemShut {NoStop}%
\bibitem [{\citenamefont {Biele}\ \emph {et~al.}(2015)\citenamefont {Biele}, \citenamefont {D’Agosta},\ and\ \citenamefont {Rubio}}]{biele2015time}%
  \BibitemOpen
  \bibfield  {author} {\bibinfo {author} {\bibfnamefont {R.}~\bibnamefont {Biele}}, \bibinfo {author} {\bibfnamefont {R.}~\bibnamefont {D’Agosta}}, \ and\ \bibinfo {author} {\bibfnamefont {A.}~\bibnamefont {Rubio}},\ }\href@noop {} {\bibfield  {journal} {\bibinfo  {journal} {Physical Review Letters}\ }\textbf {\bibinfo {volume} {115}},\ \bibinfo {pages} {056801} (\bibinfo {year} {2015})}\BibitemShut {NoStop}%
\bibitem [{\citenamefont {Allen}\ and\ \citenamefont {Feldman}(1993)}]{Allen_Feldman_PhysRevB.48.12581}%
  \BibitemOpen
  \bibfield  {author} {\bibinfo {author} {\bibfnamefont {P.~B.}\ \bibnamefont {Allen}}\ and\ \bibinfo {author} {\bibfnamefont {J.~L.}\ \bibnamefont {Feldman}},\ }\href {\doibase 10.1103/PhysRevB.48.12581} {\bibfield  {journal} {\bibinfo  {journal} {Phys. Rev. B}\ }\textbf {\bibinfo {volume} {48}},\ \bibinfo {pages} {12581} (\bibinfo {year} {1993})}\BibitemShut {NoStop}%
\bibitem [{\citenamefont {Hua}\ \emph {et~al.}(2019)\citenamefont {Hua}, \citenamefont {Lindsay}, \citenamefont {Chen},\ and\ \citenamefont {Minnich}}]{hua2019generalized}%
  \BibitemOpen
  \bibfield  {author} {\bibinfo {author} {\bibfnamefont {C.}~\bibnamefont {Hua}}, \bibinfo {author} {\bibfnamefont {L.}~\bibnamefont {Lindsay}}, \bibinfo {author} {\bibfnamefont {X.}~\bibnamefont {Chen}}, \ and\ \bibinfo {author} {\bibfnamefont {A.~J.}\ \bibnamefont {Minnich}},\ }\href@noop {} {\bibfield  {journal} {\bibinfo  {journal} {Physical Review B}\ }\textbf {\bibinfo {volume} {100}},\ \bibinfo {pages} {085203} (\bibinfo {year} {2019})}\BibitemShut {NoStop}%
\bibitem [{\citenamefont {Chen}(2021)}]{chen2021non}%
  \BibitemOpen
  \bibfield  {author} {\bibinfo {author} {\bibfnamefont {G.}~\bibnamefont {Chen}},\ }\href@noop {} {\bibfield  {journal} {\bibinfo  {journal} {Nature Reviews Physics}\ }\textbf {\bibinfo {volume} {3}},\ \bibinfo {pages} {555} (\bibinfo {year} {2021})}\BibitemShut {NoStop}%
\bibitem [{\citenamefont {Simoncelli}\ \emph {et~al.}(2019)\citenamefont {Simoncelli}, \citenamefont {Marzari},\ and\ \citenamefont {Mauri}}]{simoncelli2019unified}%
  \BibitemOpen
  \bibfield  {author} {\bibinfo {author} {\bibfnamefont {M.}~\bibnamefont {Simoncelli}}, \bibinfo {author} {\bibfnamefont {N.}~\bibnamefont {Marzari}}, \ and\ \bibinfo {author} {\bibfnamefont {F.}~\bibnamefont {Mauri}},\ }\href@noop {} {\bibfield  {journal} {\bibinfo  {journal} {Nature Physics}\ }\textbf {\bibinfo {volume} {15}},\ \bibinfo {pages} {809} (\bibinfo {year} {2019})}\BibitemShut {NoStop}%
\bibitem [{\citenamefont {Simoncelli}\ \emph {et~al.}(2020)\citenamefont {Simoncelli}, \citenamefont {Marzari},\ and\ \citenamefont {Cepellotti}}]{simoncelli2020generalization}%
  \BibitemOpen
  \bibfield  {author} {\bibinfo {author} {\bibfnamefont {M.}~\bibnamefont {Simoncelli}}, \bibinfo {author} {\bibfnamefont {N.}~\bibnamefont {Marzari}}, \ and\ \bibinfo {author} {\bibfnamefont {A.}~\bibnamefont {Cepellotti}},\ }\href@noop {} {\bibfield  {journal} {\bibinfo  {journal} {Physical Review X}\ }\textbf {\bibinfo {volume} {10}},\ \bibinfo {pages} {011019} (\bibinfo {year} {2020})}\BibitemShut {NoStop}%
\bibitem [{\citenamefont {Zeng}\ \emph {et~al.}(2021{\natexlab{a}})\citenamefont {Zeng}, \citenamefont {Avritte},\ and\ \citenamefont {Dong}}]{Zeng2021ITC}%
  \BibitemOpen
  \bibfield  {author} {\bibinfo {author} {\bibfnamefont {Y.}~\bibnamefont {Zeng}}, \bibinfo {author} {\bibfnamefont {J.~T.}\ \bibnamefont {Avritte}}, \ and\ \bibinfo {author} {\bibfnamefont {J.}~\bibnamefont {Dong}},\ }\href {\doibase 10.1002/pssb.202000454} {\bibfield  {journal} {\bibinfo  {journal} {physica status solidi (b)}\ }\textbf {\bibinfo {volume} {258}},\ \bibinfo {pages} {2000454} (\bibinfo {year} {2021}{\natexlab{a}})}\BibitemShut {NoStop}%
\bibitem [{\citenamefont {Allen}\ and\ \citenamefont {Nghiem}(2022)}]{allen2022heat}%
  \BibitemOpen
  \bibfield  {author} {\bibinfo {author} {\bibfnamefont {P.~B.}\ \bibnamefont {Allen}}\ and\ \bibinfo {author} {\bibfnamefont {N.~A.}\ \bibnamefont {Nghiem}},\ }\href@noop {} {\bibfield  {journal} {\bibinfo  {journal} {Physical Review B}\ }\textbf {\bibinfo {volume} {105}},\ \bibinfo {pages} {174302} (\bibinfo {year} {2022})}\BibitemShut {NoStop}%
\bibitem [{\citenamefont {Zeng}\ and\ \citenamefont {Khodadadi}(2018)}]{zeng2018molecular}%
  \BibitemOpen
  \bibfield  {author} {\bibinfo {author} {\bibfnamefont {Y.}~\bibnamefont {Zeng}}\ and\ \bibinfo {author} {\bibfnamefont {J.}~\bibnamefont {Khodadadi}},\ }\href@noop {} {\bibfield  {journal} {\bibinfo  {journal} {Energy \& Fuels}\ }\textbf {\bibinfo {volume} {32}},\ \bibinfo {pages} {11253} (\bibinfo {year} {2018})}\BibitemShut {NoStop}%
\bibitem [{\citenamefont {Zeng}\ \emph {et~al.}(2021{\natexlab{b}})\citenamefont {Zeng}, \citenamefont {Dong},\ and\ \citenamefont {Khodadadi}}]{zeng2021thermal}%
  \BibitemOpen
  \bibfield  {author} {\bibinfo {author} {\bibfnamefont {Y.}~\bibnamefont {Zeng}}, \bibinfo {author} {\bibfnamefont {J.}~\bibnamefont {Dong}}, \ and\ \bibinfo {author} {\bibfnamefont {J.}~\bibnamefont {Khodadadi}},\ }\href@noop {} {\bibfield  {journal} {\bibinfo  {journal} {International Journal of Heat and Mass Transfer}\ }\textbf {\bibinfo {volume} {164}},\ \bibinfo {pages} {120603} (\bibinfo {year} {2021}{\natexlab{b}})}\BibitemShut {NoStop}%
\bibitem [{\citenamefont {Dong}\ \emph {et~al.}(2001)\citenamefont {Dong}, \citenamefont {Sankey},\ and\ \citenamefont {Myles}}]{dong2001_PhysRevLett.86.2361}%
  \BibitemOpen
  \bibfield  {author} {\bibinfo {author} {\bibfnamefont {J.}~\bibnamefont {Dong}}, \bibinfo {author} {\bibfnamefont {O.~F.}\ \bibnamefont {Sankey}}, \ and\ \bibinfo {author} {\bibfnamefont {C.~W.}\ \bibnamefont {Myles}},\ }\href {\doibase 10.1103/PhysRevLett.86.2361} {\bibfield  {journal} {\bibinfo  {journal} {Phys. Rev. Lett.}\ }\textbf {\bibinfo {volume} {86}},\ \bibinfo {pages} {2361} (\bibinfo {year} {2001})}\BibitemShut {NoStop}%
\bibitem [{\citenamefont {Cahill}\ \emph {et~al.}(2003)\citenamefont {Cahill}, \citenamefont {Ford}, \citenamefont {Goodson}, \citenamefont {Mahan}, \citenamefont {Majumdar}, \citenamefont {Maris}, \citenamefont {Merlin},\ and\ \citenamefont {Phillpot}}]{cahill2003nanoscale}%
  \BibitemOpen
  \bibfield  {author} {\bibinfo {author} {\bibfnamefont {D.~G.}\ \bibnamefont {Cahill}}, \bibinfo {author} {\bibfnamefont {W.~K.}\ \bibnamefont {Ford}}, \bibinfo {author} {\bibfnamefont {K.~E.}\ \bibnamefont {Goodson}}, \bibinfo {author} {\bibfnamefont {G.~D.}\ \bibnamefont {Mahan}}, \bibinfo {author} {\bibfnamefont {A.}~\bibnamefont {Majumdar}}, \bibinfo {author} {\bibfnamefont {H.~J.}\ \bibnamefont {Maris}}, \bibinfo {author} {\bibfnamefont {R.}~\bibnamefont {Merlin}}, \ and\ \bibinfo {author} {\bibfnamefont {S.~R.}\ \bibnamefont {Phillpot}},\ }\href@noop {} {\bibfield  {journal} {\bibinfo  {journal} {Journal of applied physics}\ }\textbf {\bibinfo {volume} {93}},\ \bibinfo {pages} {793} (\bibinfo {year} {2003})}\BibitemShut {NoStop}%
\bibitem [{\citenamefont {Cahill}\ \emph {et~al.}(2014)\citenamefont {Cahill}, \citenamefont {Braun}, \citenamefont {Chen}, \citenamefont {Clarke}, \citenamefont {Fan}, \citenamefont {Goodson}, \citenamefont {Keblinski}, \citenamefont {King}, \citenamefont {Mahan}, \citenamefont {Majumdar} \emph {et~al.}}]{cahill2014nanoscale}%
  \BibitemOpen
  \bibfield  {author} {\bibinfo {author} {\bibfnamefont {D.~G.}\ \bibnamefont {Cahill}}, \bibinfo {author} {\bibfnamefont {P.~V.}\ \bibnamefont {Braun}}, \bibinfo {author} {\bibfnamefont {G.}~\bibnamefont {Chen}}, \bibinfo {author} {\bibfnamefont {D.~R.}\ \bibnamefont {Clarke}}, \bibinfo {author} {\bibfnamefont {S.}~\bibnamefont {Fan}}, \bibinfo {author} {\bibfnamefont {K.~E.}\ \bibnamefont {Goodson}}, \bibinfo {author} {\bibfnamefont {P.}~\bibnamefont {Keblinski}}, \bibinfo {author} {\bibfnamefont {W.~P.}\ \bibnamefont {King}}, \bibinfo {author} {\bibfnamefont {G.~D.}\ \bibnamefont {Mahan}}, \bibinfo {author} {\bibfnamefont {A.}~\bibnamefont {Majumdar}},  \emph {et~al.},\ }\href@noop {} {\bibfield  {journal} {\bibinfo  {journal} {Applied physics reviews}\ }\textbf {\bibinfo {volume} {1}} (\bibinfo {year} {2014})}\BibitemShut {NoStop}%
\bibitem [{\citenamefont {Koh}\ and\ \citenamefont {Cahill}(2007)}]{koh2007_PhysRevB.76.075207}%
  \BibitemOpen
  \bibfield  {author} {\bibinfo {author} {\bibfnamefont {Y.~K.}\ \bibnamefont {Koh}}\ and\ \bibinfo {author} {\bibfnamefont {D.~G.}\ \bibnamefont {Cahill}},\ }\href {\doibase 10.1103/PhysRevB.76.075207} {\bibfield  {journal} {\bibinfo  {journal} {Phys. Rev. B}\ }\textbf {\bibinfo {volume} {76}},\ \bibinfo {pages} {075207} (\bibinfo {year} {2007})}\BibitemShut {NoStop}%
\bibitem [{\citenamefont {Siemens}\ \emph {et~al.}(2010)\citenamefont {Siemens}, \citenamefont {Li}, \citenamefont {Yang}, \citenamefont {Nelson}, \citenamefont {Anderson}, \citenamefont {Murnane},\ and\ \citenamefont {Kapteyn}}]{siemens2010quasi}%
  \BibitemOpen
  \bibfield  {author} {\bibinfo {author} {\bibfnamefont {M.~E.}\ \bibnamefont {Siemens}}, \bibinfo {author} {\bibfnamefont {Q.}~\bibnamefont {Li}}, \bibinfo {author} {\bibfnamefont {R.}~\bibnamefont {Yang}}, \bibinfo {author} {\bibfnamefont {K.~A.}\ \bibnamefont {Nelson}}, \bibinfo {author} {\bibfnamefont {E.~H.}\ \bibnamefont {Anderson}}, \bibinfo {author} {\bibfnamefont {M.~M.}\ \bibnamefont {Murnane}}, \ and\ \bibinfo {author} {\bibfnamefont {H.~C.}\ \bibnamefont {Kapteyn}},\ }\href@noop {} {\bibfield  {journal} {\bibinfo  {journal} {Nature materials}\ }\textbf {\bibinfo {volume} {9}},\ \bibinfo {pages} {26} (\bibinfo {year} {2010})}\BibitemShut {NoStop}%
\bibitem [{\citenamefont {Minnich}\ \emph {et~al.}(2011)\citenamefont {Minnich}, \citenamefont {Johnson}, \citenamefont {Schmidt}, \citenamefont {Esfarjani}, \citenamefont {Dresselhaus}, \citenamefont {Nelson},\ and\ \citenamefont {Chen}}]{minnich2011thermal}%
  \BibitemOpen
  \bibfield  {author} {\bibinfo {author} {\bibfnamefont {A.~J.}\ \bibnamefont {Minnich}}, \bibinfo {author} {\bibfnamefont {J.~A.}\ \bibnamefont {Johnson}}, \bibinfo {author} {\bibfnamefont {A.~J.}\ \bibnamefont {Schmidt}}, \bibinfo {author} {\bibfnamefont {K.}~\bibnamefont {Esfarjani}}, \bibinfo {author} {\bibfnamefont {M.~S.}\ \bibnamefont {Dresselhaus}}, \bibinfo {author} {\bibfnamefont {K.~A.}\ \bibnamefont {Nelson}}, \ and\ \bibinfo {author} {\bibfnamefont {G.}~\bibnamefont {Chen}},\ }\href@noop {} {\bibfield  {journal} {\bibinfo  {journal} {Physical review letters}\ }\textbf {\bibinfo {volume} {107}},\ \bibinfo {pages} {095901} (\bibinfo {year} {2011})}\BibitemShut {NoStop}%
\bibitem [{\citenamefont {Johnson}\ \emph {et~al.}(2013)\citenamefont {Johnson}, \citenamefont {Maznev}, \citenamefont {Cuffe}, \citenamefont {Eliason}, \citenamefont {Minnich}, \citenamefont {Kehoe}, \citenamefont {Torres}, \citenamefont {Chen},\ and\ \citenamefont {Nelson}}]{johnson2013direct}%
  \BibitemOpen
  \bibfield  {author} {\bibinfo {author} {\bibfnamefont {J.~A.}\ \bibnamefont {Johnson}}, \bibinfo {author} {\bibfnamefont {A.}~\bibnamefont {Maznev}}, \bibinfo {author} {\bibfnamefont {J.}~\bibnamefont {Cuffe}}, \bibinfo {author} {\bibfnamefont {J.~K.}\ \bibnamefont {Eliason}}, \bibinfo {author} {\bibfnamefont {A.~J.}\ \bibnamefont {Minnich}}, \bibinfo {author} {\bibfnamefont {T.}~\bibnamefont {Kehoe}}, \bibinfo {author} {\bibfnamefont {C.~M.~S.}\ \bibnamefont {Torres}}, \bibinfo {author} {\bibfnamefont {G.}~\bibnamefont {Chen}}, \ and\ \bibinfo {author} {\bibfnamefont {K.~A.}\ \bibnamefont {Nelson}},\ }\href@noop {} {\bibfield  {journal} {\bibinfo  {journal} {Physical Review Letters}\ }\textbf {\bibinfo {volume} {110}},\ \bibinfo {pages} {025901} (\bibinfo {year} {2013})}\BibitemShut {NoStop}%
\bibitem [{\citenamefont {Hu}\ \emph {et~al.}(2015)\citenamefont {Hu}, \citenamefont {Zeng}, \citenamefont {Minnich}, \citenamefont {Dresselhaus},\ and\ \citenamefont {Chen}}]{Hu2015SpectralMapping}%
  \BibitemOpen
  \bibfield  {author} {\bibinfo {author} {\bibfnamefont {Y.}~\bibnamefont {Hu}}, \bibinfo {author} {\bibfnamefont {L.}~\bibnamefont {Zeng}}, \bibinfo {author} {\bibfnamefont {A.~J.}\ \bibnamefont {Minnich}}, \bibinfo {author} {\bibfnamefont {M.~S.}\ \bibnamefont {Dresselhaus}}, \ and\ \bibinfo {author} {\bibfnamefont {G.}~\bibnamefont {Chen}},\ }\href {\doibase 10.1038/nnano.2015.109} {\bibfield  {journal} {\bibinfo  {journal} {Nature Nanotechnology}\ }\textbf {\bibinfo {volume} {10}},\ \bibinfo {pages} {701} (\bibinfo {year} {2015})}\BibitemShut {NoStop}%
\bibitem [{\citenamefont {Hoogeboom-Pot}\ \emph {et~al.}(2015)\citenamefont {Hoogeboom-Pot}, \citenamefont {Hernandez-Charpak}, \citenamefont {Frazer}, \citenamefont {Gu}, \citenamefont {Turgut}, \citenamefont {Anderson}, \citenamefont {Chao}, \citenamefont {Shaw}, \citenamefont {Yang}, \citenamefont {Murnane} \emph {et~al.}}]{hoogeboom2015mechanical}%
  \BibitemOpen
  \bibfield  {author} {\bibinfo {author} {\bibfnamefont {K.}~\bibnamefont {Hoogeboom-Pot}}, \bibinfo {author} {\bibfnamefont {J.}~\bibnamefont {Hernandez-Charpak}}, \bibinfo {author} {\bibfnamefont {T.}~\bibnamefont {Frazer}}, \bibinfo {author} {\bibfnamefont {X.}~\bibnamefont {Gu}}, \bibinfo {author} {\bibfnamefont {E.}~\bibnamefont {Turgut}}, \bibinfo {author} {\bibfnamefont {E.}~\bibnamefont {Anderson}}, \bibinfo {author} {\bibfnamefont {W.}~\bibnamefont {Chao}}, \bibinfo {author} {\bibfnamefont {J.}~\bibnamefont {Shaw}}, \bibinfo {author} {\bibfnamefont {R.}~\bibnamefont {Yang}}, \bibinfo {author} {\bibfnamefont {M.}~\bibnamefont {Murnane}},  \emph {et~al.},\ }in\ \href@noop {} {\emph {\bibinfo {booktitle} {Metrology, Inspection, and Process Control for Microlithography XXIX}}},\ Vol.\ \bibinfo {volume} {9424}\ (\bibinfo {organization} {SPIE},\ \bibinfo {year} {2015})\ pp.\ \bibinfo {pages} {401--408}\BibitemShut {NoStop}%
\bibitem [{\citenamefont {Frazer}\ \emph {et~al.}(2019)\citenamefont {Frazer}, \citenamefont {Knobloch}, \citenamefont {Hoogeboom-Pot}, \citenamefont {Nardi}, \citenamefont {Chao}, \citenamefont {Falcone}, \citenamefont {Murnane}, \citenamefont {Kapteyn},\ and\ \citenamefont {Hernandez-Charpak}}]{Frazer2019_PhysRevApplied.11.024042}%
  \BibitemOpen
  \bibfield  {author} {\bibinfo {author} {\bibfnamefont {T.~D.}\ \bibnamefont {Frazer}}, \bibinfo {author} {\bibfnamefont {J.~L.}\ \bibnamefont {Knobloch}}, \bibinfo {author} {\bibfnamefont {K.~M.}\ \bibnamefont {Hoogeboom-Pot}}, \bibinfo {author} {\bibfnamefont {D.}~\bibnamefont {Nardi}}, \bibinfo {author} {\bibfnamefont {W.}~\bibnamefont {Chao}}, \bibinfo {author} {\bibfnamefont {R.~W.}\ \bibnamefont {Falcone}}, \bibinfo {author} {\bibfnamefont {M.~M.}\ \bibnamefont {Murnane}}, \bibinfo {author} {\bibfnamefont {H.~C.}\ \bibnamefont {Kapteyn}}, \ and\ \bibinfo {author} {\bibfnamefont {J.~N.}\ \bibnamefont {Hernandez-Charpak}},\ }\href {\doibase 10.1103/PhysRevApplied.11.024042} {\bibfield  {journal} {\bibinfo  {journal} {Phys. Rev. Appl.}\ }\textbf {\bibinfo {volume} {11}},\ \bibinfo {pages} {024042} (\bibinfo {year} {2019})}\BibitemShut {NoStop}%
\bibitem [{\citenamefont {Bencivenga}\ \emph {et~al.}(2019)\citenamefont {Bencivenga}, \citenamefont {Mincigrucci}, \citenamefont {Capotondi}, \citenamefont {Foglia}, \citenamefont {Naumenko}, \citenamefont {Maznev}, \citenamefont {Pedersoli}, \citenamefont {Simoncig}, \citenamefont {Caporaletti}, \citenamefont {Chiloyan} \emph {et~al.}}]{2019_EUV_TTG}%
  \BibitemOpen
  \bibfield  {author} {\bibinfo {author} {\bibfnamefont {F.}~\bibnamefont {Bencivenga}}, \bibinfo {author} {\bibfnamefont {R.}~\bibnamefont {Mincigrucci}}, \bibinfo {author} {\bibfnamefont {F.}~\bibnamefont {Capotondi}}, \bibinfo {author} {\bibfnamefont {L.}~\bibnamefont {Foglia}}, \bibinfo {author} {\bibfnamefont {D.}~\bibnamefont {Naumenko}}, \bibinfo {author} {\bibfnamefont {A.}~\bibnamefont {Maznev}}, \bibinfo {author} {\bibfnamefont {E.}~\bibnamefont {Pedersoli}}, \bibinfo {author} {\bibfnamefont {A.}~\bibnamefont {Simoncig}}, \bibinfo {author} {\bibfnamefont {F.}~\bibnamefont {Caporaletti}}, \bibinfo {author} {\bibfnamefont {V.}~\bibnamefont {Chiloyan}},  \emph {et~al.},\ }\href@noop {} {\bibfield  {journal} {\bibinfo  {journal} {Science advances}\ }\textbf {\bibinfo {volume} {5}},\ \bibinfo {pages} {eaaw5805} (\bibinfo {year} {2019})}\BibitemShut {NoStop}%
\bibitem [{\citenamefont {Beardo}\ \emph {et~al.}(2021{\natexlab{a}})\citenamefont {Beardo}, \citenamefont {Knobloch}, \citenamefont {Sendra}, \citenamefont {Bafaluy}, \citenamefont {Frazer}, \citenamefont {Chao}, \citenamefont {Hernandez-Charpak}, \citenamefont {Kapteyn}, \citenamefont {Abad}, \citenamefont {Murnane} \emph {et~al.}}]{beardo2021general}%
  \BibitemOpen
  \bibfield  {author} {\bibinfo {author} {\bibfnamefont {A.}~\bibnamefont {Beardo}}, \bibinfo {author} {\bibfnamefont {J.~L.}\ \bibnamefont {Knobloch}}, \bibinfo {author} {\bibfnamefont {L.}~\bibnamefont {Sendra}}, \bibinfo {author} {\bibfnamefont {J.}~\bibnamefont {Bafaluy}}, \bibinfo {author} {\bibfnamefont {T.~D.}\ \bibnamefont {Frazer}}, \bibinfo {author} {\bibfnamefont {W.}~\bibnamefont {Chao}}, \bibinfo {author} {\bibfnamefont {J.~N.}\ \bibnamefont {Hernandez-Charpak}}, \bibinfo {author} {\bibfnamefont {H.~C.}\ \bibnamefont {Kapteyn}}, \bibinfo {author} {\bibfnamefont {B.}~\bibnamefont {Abad}}, \bibinfo {author} {\bibfnamefont {M.~M.}\ \bibnamefont {Murnane}},  \emph {et~al.},\ }\href@noop {} {\bibfield  {journal} {\bibinfo  {journal} {ACS nano}\ }\textbf {\bibinfo {volume} {15}},\ \bibinfo {pages} {13019} (\bibinfo {year} {2021}{\natexlab{a}})}\BibitemShut {NoStop}%
\bibitem [{\citenamefont {Jeong}\ \emph {et~al.}(2021)\citenamefont {Jeong}, \citenamefont {Li}, \citenamefont {Lee}, \citenamefont {Shi},\ and\ \citenamefont {Wang}}]{Jeong_etal_PhysRevLett.127.085901}%
  \BibitemOpen
  \bibfield  {author} {\bibinfo {author} {\bibfnamefont {J.}~\bibnamefont {Jeong}}, \bibinfo {author} {\bibfnamefont {X.}~\bibnamefont {Li}}, \bibinfo {author} {\bibfnamefont {S.}~\bibnamefont {Lee}}, \bibinfo {author} {\bibfnamefont {L.}~\bibnamefont {Shi}}, \ and\ \bibinfo {author} {\bibfnamefont {Y.}~\bibnamefont {Wang}},\ }\href {\doibase 10.1103/PhysRevLett.127.085901} {\bibfield  {journal} {\bibinfo  {journal} {Phys. Rev. Lett.}\ }\textbf {\bibinfo {volume} {127}},\ \bibinfo {pages} {085901} (\bibinfo {year} {2021})}\BibitemShut {NoStop}%
\bibitem [{\citenamefont {McBennett}\ \emph {et~al.}(2023)\citenamefont {McBennett}, \citenamefont {Beardo}, \citenamefont {Nelson}, \citenamefont {Abad}, \citenamefont {Frazer}, \citenamefont {Adak}, \citenamefont {Esashi}, \citenamefont {Li}, \citenamefont {Kapteyn}, \citenamefont {Murnane} \emph {et~al.}}]{mcbennett2023universal}%
  \BibitemOpen
  \bibfield  {author} {\bibinfo {author} {\bibfnamefont {B.}~\bibnamefont {McBennett}}, \bibinfo {author} {\bibfnamefont {A.}~\bibnamefont {Beardo}}, \bibinfo {author} {\bibfnamefont {E.~E.}\ \bibnamefont {Nelson}}, \bibinfo {author} {\bibfnamefont {B.}~\bibnamefont {Abad}}, \bibinfo {author} {\bibfnamefont {T.~D.}\ \bibnamefont {Frazer}}, \bibinfo {author} {\bibfnamefont {A.}~\bibnamefont {Adak}}, \bibinfo {author} {\bibfnamefont {Y.}~\bibnamefont {Esashi}}, \bibinfo {author} {\bibfnamefont {B.}~\bibnamefont {Li}}, \bibinfo {author} {\bibfnamefont {H.~C.}\ \bibnamefont {Kapteyn}}, \bibinfo {author} {\bibfnamefont {M.~M.}\ \bibnamefont {Murnane}},  \emph {et~al.},\ }\href@noop {} {\bibfield  {journal} {\bibinfo  {journal} {Nano Letters}\ }\textbf {\bibinfo {volume} {23}},\ \bibinfo {pages} {2129} (\bibinfo {year} {2023})}\BibitemShut {NoStop}%
\bibitem [{\citenamefont {Peshkov}(1944)}]{Peshkov1944}%
  \BibitemOpen
  \bibfield  {author} {\bibinfo {author} {\bibfnamefont {V.}~\bibnamefont {Peshkov}},\ }\href@noop {} {\bibfield  {journal} {\bibinfo  {journal} {Journal of Physics (USSR)}\ }\textbf {\bibinfo {volume} {8}},\ \bibinfo {pages} {381} (\bibinfo {year} {1944})}\BibitemShut {NoStop}%
\bibitem [{\citenamefont {Ackerman}\ \emph {et~al.}(1966)\citenamefont {Ackerman}, \citenamefont {Bertman}, \citenamefont {Fairbank},\ and\ \citenamefont {Guyer}}]{ackerman1966second}%
  \BibitemOpen
  \bibfield  {author} {\bibinfo {author} {\bibfnamefont {C.~C.}\ \bibnamefont {Ackerman}}, \bibinfo {author} {\bibfnamefont {B.}~\bibnamefont {Bertman}}, \bibinfo {author} {\bibfnamefont {H.~A.}\ \bibnamefont {Fairbank}}, \ and\ \bibinfo {author} {\bibfnamefont {R.}~\bibnamefont {Guyer}},\ }\href@noop {} {\bibfield  {journal} {\bibinfo  {journal} {Physical Review Letters}\ }\textbf {\bibinfo {volume} {16}},\ \bibinfo {pages} {789} (\bibinfo {year} {1966})}\BibitemShut {NoStop}%
\bibitem [{\citenamefont {Jackson}\ \emph {et~al.}(1970)\citenamefont {Jackson}, \citenamefont {Walker},\ and\ \citenamefont {McNelly}}]{jackson1970second}%
  \BibitemOpen
  \bibfield  {author} {\bibinfo {author} {\bibfnamefont {H.~E.}\ \bibnamefont {Jackson}}, \bibinfo {author} {\bibfnamefont {C.~T.}\ \bibnamefont {Walker}}, \ and\ \bibinfo {author} {\bibfnamefont {T.~F.}\ \bibnamefont {McNelly}},\ }\href@noop {} {\bibfield  {journal} {\bibinfo  {journal} {Physical Review Letters}\ }\textbf {\bibinfo {volume} {25}},\ \bibinfo {pages} {26} (\bibinfo {year} {1970})}\BibitemShut {NoStop}%
\bibitem [{\citenamefont {Narayanamurti}\ and\ \citenamefont {Dynes}(1972)}]{narayanamurti1972observation}%
  \BibitemOpen
  \bibfield  {author} {\bibinfo {author} {\bibfnamefont {V.}~\bibnamefont {Narayanamurti}}\ and\ \bibinfo {author} {\bibfnamefont {R.}~\bibnamefont {Dynes}},\ }\href@noop {} {\bibfield  {journal} {\bibinfo  {journal} {Physical Review Letters}\ }\textbf {\bibinfo {volume} {28}},\ \bibinfo {pages} {1461} (\bibinfo {year} {1972})}\BibitemShut {NoStop}%
\bibitem [{\citenamefont {Koreeda}\ \emph {et~al.}(2007)\citenamefont {Koreeda}, \citenamefont {Takano},\ and\ \citenamefont {Saikan}}]{koreeda2007second}%
  \BibitemOpen
  \bibfield  {author} {\bibinfo {author} {\bibfnamefont {A.}~\bibnamefont {Koreeda}}, \bibinfo {author} {\bibfnamefont {R.}~\bibnamefont {Takano}}, \ and\ \bibinfo {author} {\bibfnamefont {S.}~\bibnamefont {Saikan}},\ }\href@noop {} {\bibfield  {journal} {\bibinfo  {journal} {Physical Review Letters}\ }\textbf {\bibinfo {volume} {99}},\ \bibinfo {pages} {265502} (\bibinfo {year} {2007})}\BibitemShut {NoStop}%
\bibitem [{\citenamefont {Beardo}\ \emph {et~al.}(2021{\natexlab{b}})\citenamefont {Beardo}, \citenamefont {L{\'o}pez-Su{\'a}rez}, \citenamefont {P{\'e}rez}, \citenamefont {Sendra}, \citenamefont {Alonso}, \citenamefont {Melis}, \citenamefont {Bafaluy}, \citenamefont {Camacho}, \citenamefont {Colombo}, \citenamefont {Rurali} \emph {et~al.}}]{beardo2021observation}%
  \BibitemOpen
  \bibfield  {author} {\bibinfo {author} {\bibfnamefont {A.}~\bibnamefont {Beardo}}, \bibinfo {author} {\bibfnamefont {M.}~\bibnamefont {L{\'o}pez-Su{\'a}rez}}, \bibinfo {author} {\bibfnamefont {L.~A.}\ \bibnamefont {P{\'e}rez}}, \bibinfo {author} {\bibfnamefont {L.}~\bibnamefont {Sendra}}, \bibinfo {author} {\bibfnamefont {M.~I.}\ \bibnamefont {Alonso}}, \bibinfo {author} {\bibfnamefont {C.}~\bibnamefont {Melis}}, \bibinfo {author} {\bibfnamefont {J.}~\bibnamefont {Bafaluy}}, \bibinfo {author} {\bibfnamefont {J.}~\bibnamefont {Camacho}}, \bibinfo {author} {\bibfnamefont {L.}~\bibnamefont {Colombo}}, \bibinfo {author} {\bibfnamefont {R.}~\bibnamefont {Rurali}},  \emph {et~al.},\ }\href@noop {} {\bibfield  {journal} {\bibinfo  {journal} {Science advances}\ }\textbf {\bibinfo {volume} {7}},\ \bibinfo {pages} {eabg4677} (\bibinfo {year} {2021}{\natexlab{b}})}\BibitemShut {NoStop}%
\bibitem [{\citenamefont {Huberman}\ \emph {et~al.}(2019)\citenamefont {Huberman}, \citenamefont {Duncan}, \citenamefont {Chen}, \citenamefont {Song}, \citenamefont {Chiloyan}, \citenamefont {Ding}, \citenamefont {Maznev}, \citenamefont {Chen},\ and\ \citenamefont {Nelson}}]{huberman2019observation}%
  \BibitemOpen
  \bibfield  {author} {\bibinfo {author} {\bibfnamefont {S.}~\bibnamefont {Huberman}}, \bibinfo {author} {\bibfnamefont {R.~A.}\ \bibnamefont {Duncan}}, \bibinfo {author} {\bibfnamefont {K.}~\bibnamefont {Chen}}, \bibinfo {author} {\bibfnamefont {B.}~\bibnamefont {Song}}, \bibinfo {author} {\bibfnamefont {V.}~\bibnamefont {Chiloyan}}, \bibinfo {author} {\bibfnamefont {Z.}~\bibnamefont {Ding}}, \bibinfo {author} {\bibfnamefont {A.~A.}\ \bibnamefont {Maznev}}, \bibinfo {author} {\bibfnamefont {G.}~\bibnamefont {Chen}}, \ and\ \bibinfo {author} {\bibfnamefont {K.~A.}\ \bibnamefont {Nelson}},\ }\href@noop {} {\bibfield  {journal} {\bibinfo  {journal} {Science}\ }\textbf {\bibinfo {volume} {364}},\ \bibinfo {pages} {375} (\bibinfo {year} {2019})}\BibitemShut {NoStop}%
\bibitem [{\citenamefont {Ding}\ \emph {et~al.}(2022)\citenamefont {Ding}, \citenamefont {Chen}, \citenamefont {Song}, \citenamefont {Shin}, \citenamefont {Maznev}, \citenamefont {Nelson},\ and\ \citenamefont {Chen}}]{ding2022observation}%
  \BibitemOpen
  \bibfield  {author} {\bibinfo {author} {\bibfnamefont {Z.}~\bibnamefont {Ding}}, \bibinfo {author} {\bibfnamefont {K.}~\bibnamefont {Chen}}, \bibinfo {author} {\bibfnamefont {B.}~\bibnamefont {Song}}, \bibinfo {author} {\bibfnamefont {J.}~\bibnamefont {Shin}}, \bibinfo {author} {\bibfnamefont {A.~A.}\ \bibnamefont {Maznev}}, \bibinfo {author} {\bibfnamefont {K.~A.}\ \bibnamefont {Nelson}}, \ and\ \bibinfo {author} {\bibfnamefont {G.}~\bibnamefont {Chen}},\ }\href@noop {} {\bibfield  {journal} {\bibinfo  {journal} {Nature communications}\ }\textbf {\bibinfo {volume} {13}},\ \bibinfo {pages} {285} (\bibinfo {year} {2022})}\BibitemShut {NoStop}%
\bibitem [{\citenamefont {Joseph}\ and\ \citenamefont {Preziosi}(1989)}]{joseph1989heat}%
  \BibitemOpen
  \bibfield  {author} {\bibinfo {author} {\bibfnamefont {D.~D.}\ \bibnamefont {Joseph}}\ and\ \bibinfo {author} {\bibfnamefont {L.}~\bibnamefont {Preziosi}},\ }\href@noop {} {\bibfield  {journal} {\bibinfo  {journal} {Reviews of modern physics}\ }\textbf {\bibinfo {volume} {61}},\ \bibinfo {pages} {41} (\bibinfo {year} {1989})}\BibitemShut {NoStop}%
\bibitem [{\citenamefont {Maxwell}(2003)}]{maxwell1866dynamical}%
  \BibitemOpen
  \bibfield  {author} {\bibinfo {author} {\bibfnamefont {J.~C.}\ \bibnamefont {Maxwell}},\ }in\ \href@noop {} {\emph {\bibinfo {booktitle} {The kinetic theory of gases: an anthology of classic papers with historical commentary}}}\ (\bibinfo  {publisher} {World Scientific},\ \bibinfo {year} {2003})\ pp.\ \bibinfo {pages} {197--261}\BibitemShut {NoStop}%
\bibitem [{\citenamefont {Cattaneo}(1948)}]{cattaneo1948atti}%
  \BibitemOpen
  \bibfield  {author} {\bibinfo {author} {\bibfnamefont {C.}~\bibnamefont {Cattaneo}},\ }\href@noop {} {\bibfield  {journal} {\bibinfo  {journal} {CR Acad. Sci. Paris}\ }\textbf {\bibinfo {volume} {247}},\ \bibinfo {pages} {431} (\bibinfo {year} {1948})}\BibitemShut {NoStop}%
\bibitem [{\citenamefont {Vernotte}(1958)}]{vernotte1958paradoxes}%
  \BibitemOpen
  \bibfield  {author} {\bibinfo {author} {\bibfnamefont {P.}~\bibnamefont {Vernotte}},\ }\href@noop {} {\bibfield  {journal} {\bibinfo  {journal} {Comptes rendus}\ }\textbf {\bibinfo {volume} {246}},\ \bibinfo {pages} {3154} (\bibinfo {year} {1958})}\BibitemShut {NoStop}%
\bibitem [{\citenamefont {Dong}\ \emph {et~al.}(1999)\citenamefont {Dong}, \citenamefont {Sankey},\ and\ \citenamefont {Kern}}]{PhysRevB.60.950}%
  \BibitemOpen
  \bibfield  {author} {\bibinfo {author} {\bibfnamefont {J.}~\bibnamefont {Dong}}, \bibinfo {author} {\bibfnamefont {O.~F.}\ \bibnamefont {Sankey}}, \ and\ \bibinfo {author} {\bibfnamefont {G.}~\bibnamefont {Kern}},\ }\href {\doibase 10.1103/PhysRevB.60.950} {\bibfield  {journal} {\bibinfo  {journal} {Phys. Rev. B}\ }\textbf {\bibinfo {volume} {60}},\ \bibinfo {pages} {950} (\bibinfo {year} {1999})}\BibitemShut {NoStop}%
\bibitem [{\citenamefont {Tang}\ \emph {et~al.}(2006)\citenamefont {Tang}, \citenamefont {Dong}, \citenamefont {Hutchins}, \citenamefont {Shebanova}, \citenamefont {Gryko}, \citenamefont {Barnes}, \citenamefont {Cockroft}, \citenamefont {Vickers},\ and\ \citenamefont {McMillan}}]{PhysRevB.74.014109}%
  \BibitemOpen
  \bibfield  {author} {\bibinfo {author} {\bibfnamefont {X.}~\bibnamefont {Tang}}, \bibinfo {author} {\bibfnamefont {J.}~\bibnamefont {Dong}}, \bibinfo {author} {\bibfnamefont {P.}~\bibnamefont {Hutchins}}, \bibinfo {author} {\bibfnamefont {O.}~\bibnamefont {Shebanova}}, \bibinfo {author} {\bibfnamefont {J.}~\bibnamefont {Gryko}}, \bibinfo {author} {\bibfnamefont {P.}~\bibnamefont {Barnes}}, \bibinfo {author} {\bibfnamefont {J.~K.}\ \bibnamefont {Cockroft}}, \bibinfo {author} {\bibfnamefont {M.}~\bibnamefont {Vickers}}, \ and\ \bibinfo {author} {\bibfnamefont {P.~F.}\ \bibnamefont {McMillan}},\ }\href {\doibase 10.1103/PhysRevB.74.014109} {\bibfield  {journal} {\bibinfo  {journal} {Phys. Rev. B}\ }\textbf {\bibinfo {volume} {74}},\ \bibinfo {pages} {014109} (\bibinfo {year} {2006})}\BibitemShut {NoStop}%
\bibitem [{\citenamefont {Ward}\ \emph {et~al.}(2009)\citenamefont {Ward}, \citenamefont {Broido}, \citenamefont {Stewart},\ and\ \citenamefont {Deinzer}}]{ward2009ab}%
  \BibitemOpen
  \bibfield  {author} {\bibinfo {author} {\bibfnamefont {A.}~\bibnamefont {Ward}}, \bibinfo {author} {\bibfnamefont {D.~A.}\ \bibnamefont {Broido}}, \bibinfo {author} {\bibfnamefont {D.~A.}\ \bibnamefont {Stewart}}, \ and\ \bibinfo {author} {\bibfnamefont {G.}~\bibnamefont {Deinzer}},\ }\href@noop {} {\bibfield  {journal} {\bibinfo  {journal} {Phys. Rev. B}\ }\textbf {\bibinfo {volume} {80}},\ \bibinfo {pages} {125203} (\bibinfo {year} {2009})}\BibitemShut {NoStop}%
\bibitem [{\citenamefont {Tang}\ and\ \citenamefont {Dong}(2009)}]{tang2009pressure}%
  \BibitemOpen
  \bibfield  {author} {\bibinfo {author} {\bibfnamefont {X.}~\bibnamefont {Tang}}\ and\ \bibinfo {author} {\bibfnamefont {J.}~\bibnamefont {Dong}},\ }\href@noop {} {\bibfield  {journal} {\bibinfo  {journal} {Phys. Eart. Planet. Interi.}\ }\textbf {\bibinfo {volume} {174}},\ \bibinfo {pages} {33} (\bibinfo {year} {2009})}\BibitemShut {NoStop}%
\bibitem [{\citenamefont {Togo}\ and\ \citenamefont {Tanaka}(2015)}]{phonopy}%
  \BibitemOpen
  \bibfield  {author} {\bibinfo {author} {\bibfnamefont {A.}~\bibnamefont {Togo}}\ and\ \bibinfo {author} {\bibfnamefont {I.}~\bibnamefont {Tanaka}},\ }\href@noop {} {\bibfield  {journal} {\bibinfo  {journal} {Scr. Mater.}\ }\textbf {\bibinfo {volume} {108}},\ \bibinfo {pages} {1} (\bibinfo {year} {2015})}\BibitemShut {NoStop}%
\bibitem [{\citenamefont {Li}\ \emph {et~al.}(2014)\citenamefont {Li}, \citenamefont {Carrete}, \citenamefont {Katcho},\ and\ \citenamefont {Mingo}}]{li2014shengbte}%
  \BibitemOpen
  \bibfield  {author} {\bibinfo {author} {\bibfnamefont {W.}~\bibnamefont {Li}}, \bibinfo {author} {\bibfnamefont {J.}~\bibnamefont {Carrete}}, \bibinfo {author} {\bibfnamefont {N.~A.}\ \bibnamefont {Katcho}}, \ and\ \bibinfo {author} {\bibfnamefont {N.}~\bibnamefont {Mingo}},\ }\href@noop {} {\bibfield  {journal} {\bibinfo  {journal} {Computer Physics Communications}\ }\textbf {\bibinfo {volume} {185}},\ \bibinfo {pages} {1747} (\bibinfo {year} {2014})}\BibitemShut {NoStop}%
\bibitem [{\citenamefont {Feng}\ and\ \citenamefont {Ruan}(2016)}]{feng2016_PhysRevB.93.045202}%
  \BibitemOpen
  \bibfield  {author} {\bibinfo {author} {\bibfnamefont {T.}~\bibnamefont {Feng}}\ and\ \bibinfo {author} {\bibfnamefont {X.}~\bibnamefont {Ruan}},\ }\href {\doibase 10.1103/PhysRevB.93.045202} {\bibfield  {journal} {\bibinfo  {journal} {Phys. Rev. B}\ }\textbf {\bibinfo {volume} {93}},\ \bibinfo {pages} {045202} (\bibinfo {year} {2016})}\BibitemShut {NoStop}%
\bibitem [{\citenamefont {Ravichandran}\ and\ \citenamefont {Broido}(2020)}]{ravi2020_PhysRevX.10.021063}%
  \BibitemOpen
  \bibfield  {author} {\bibinfo {author} {\bibfnamefont {N.~K.}\ \bibnamefont {Ravichandran}}\ and\ \bibinfo {author} {\bibfnamefont {D.}~\bibnamefont {Broido}},\ }\href {\doibase 10.1103/PhysRevX.10.021063} {\bibfield  {journal} {\bibinfo  {journal} {Phys. Rev. X}\ }\textbf {\bibinfo {volume} {10}},\ \bibinfo {pages} {021063} (\bibinfo {year} {2020})}\BibitemShut {NoStop}%
\bibitem [{\citenamefont {Tang}\ and\ \citenamefont {Dong}(2010)}]{Tang4539}%
  \BibitemOpen
  \bibfield  {author} {\bibinfo {author} {\bibfnamefont {X.}~\bibnamefont {Tang}}\ and\ \bibinfo {author} {\bibfnamefont {J.}~\bibnamefont {Dong}},\ }\href {\doibase 10.1073/pnas.0907194107} {\bibfield  {journal} {\bibinfo  {journal} {Proc. Nat. Acad. of Sci.}\ }\textbf {\bibinfo {volume} {107}},\ \bibinfo {pages} {4539} (\bibinfo {year} {2010})}\BibitemShut {NoStop}%
\bibitem [{\citenamefont {Tang}\ \emph {et~al.}(2014)\citenamefont {Tang}, \citenamefont {Ntam}, \citenamefont {Dong}, \citenamefont {Rainey},\ and\ \citenamefont {Kavner}}]{tang2014thermal}%
  \BibitemOpen
  \bibfield  {author} {\bibinfo {author} {\bibfnamefont {X.}~\bibnamefont {Tang}}, \bibinfo {author} {\bibfnamefont {M.~C.}\ \bibnamefont {Ntam}}, \bibinfo {author} {\bibfnamefont {J.}~\bibnamefont {Dong}}, \bibinfo {author} {\bibfnamefont {E.~S.}\ \bibnamefont {Rainey}}, \ and\ \bibinfo {author} {\bibfnamefont {A.}~\bibnamefont {Kavner}},\ }\href@noop {} {\bibfield  {journal} {\bibinfo  {journal} {Geophys. Res. Lett.}\ }\textbf {\bibinfo {volume} {41}},\ \bibinfo {pages} {2746} (\bibinfo {year} {2014})}\BibitemShut {NoStop}%
\bibitem [{\citenamefont {Togo}\ \emph {et~al.}(2015)\citenamefont {Togo}, \citenamefont {Chaput},\ and\ \citenamefont {Tanaka}}]{phono3py}%
  \BibitemOpen
  \bibfield  {author} {\bibinfo {author} {\bibfnamefont {A.}~\bibnamefont {Togo}}, \bibinfo {author} {\bibfnamefont {L.}~\bibnamefont {Chaput}}, \ and\ \bibinfo {author} {\bibfnamefont {I.}~\bibnamefont {Tanaka}},\ }\href {\doibase 10.1103/PhysRevB.91.094306} {\bibfield  {journal} {\bibinfo  {journal} {Phys. Rev. B}\ }\textbf {\bibinfo {volume} {91}},\ \bibinfo {pages} {094306} (\bibinfo {year} {2015})}\BibitemShut {NoStop}%
\bibitem [{\citenamefont {Hua}\ and\ \citenamefont {Minnich}(2014)}]{hua2014analytical}%
  \BibitemOpen
  \bibfield  {author} {\bibinfo {author} {\bibfnamefont {C.}~\bibnamefont {Hua}}\ and\ \bibinfo {author} {\bibfnamefont {A.~J.}\ \bibnamefont {Minnich}},\ }\href@noop {} {\bibfield  {journal} {\bibinfo  {journal} {Physical Review B}\ }\textbf {\bibinfo {volume} {90}},\ \bibinfo {pages} {214306} (\bibinfo {year} {2014})}\BibitemShut {NoStop}%
\bibitem [{\citenamefont {De~Tomas}\ \emph {et~al.}(2014)\citenamefont {De~Tomas}, \citenamefont {Cantarero}, \citenamefont {Lopeandia},\ and\ \citenamefont {Alvarez}}]{de2014thermal}%
  \BibitemOpen
  \bibfield  {author} {\bibinfo {author} {\bibfnamefont {C.}~\bibnamefont {De~Tomas}}, \bibinfo {author} {\bibfnamefont {A.}~\bibnamefont {Cantarero}}, \bibinfo {author} {\bibfnamefont {A.}~\bibnamefont {Lopeandia}}, \ and\ \bibinfo {author} {\bibfnamefont {F.}~\bibnamefont {Alvarez}},\ }\href@noop {} {\bibfield  {journal} {\bibinfo  {journal} {Proceedings of the Royal Society A: Mathematical, Physical and Engineering Sciences}\ }\textbf {\bibinfo {volume} {470}},\ \bibinfo {pages} {20140371} (\bibinfo {year} {2014})}\BibitemShut {NoStop}%
\bibitem [{\citenamefont {Lee}\ \emph {et~al.}(2015)\citenamefont {Lee}, \citenamefont {Broido}, \citenamefont {Esfarjani},\ and\ \citenamefont {Chen}}]{lee2015hydrodynamic}%
  \BibitemOpen
  \bibfield  {author} {\bibinfo {author} {\bibfnamefont {S.}~\bibnamefont {Lee}}, \bibinfo {author} {\bibfnamefont {D.}~\bibnamefont {Broido}}, \bibinfo {author} {\bibfnamefont {K.}~\bibnamefont {Esfarjani}}, \ and\ \bibinfo {author} {\bibfnamefont {G.}~\bibnamefont {Chen}},\ }\href@noop {} {\bibfield  {journal} {\bibinfo  {journal} {Nature communications}\ }\textbf {\bibinfo {volume} {6}},\ \bibinfo {pages} {6290} (\bibinfo {year} {2015})}\BibitemShut {NoStop}%
\bibitem [{\citenamefont {Cepellotti}\ \emph {et~al.}(2015)\citenamefont {Cepellotti}, \citenamefont {Fugallo}, \citenamefont {Paulatto}, \citenamefont {Lazzeri}, \citenamefont {Mauri},\ and\ \citenamefont {Marzari}}]{cepellotti2015phonon}%
  \BibitemOpen
  \bibfield  {author} {\bibinfo {author} {\bibfnamefont {A.}~\bibnamefont {Cepellotti}}, \bibinfo {author} {\bibfnamefont {G.}~\bibnamefont {Fugallo}}, \bibinfo {author} {\bibfnamefont {L.}~\bibnamefont {Paulatto}}, \bibinfo {author} {\bibfnamefont {M.}~\bibnamefont {Lazzeri}}, \bibinfo {author} {\bibfnamefont {F.}~\bibnamefont {Mauri}}, \ and\ \bibinfo {author} {\bibfnamefont {N.}~\bibnamefont {Marzari}},\ }\href@noop {} {\bibfield  {journal} {\bibinfo  {journal} {Nature communications}\ }\textbf {\bibinfo {volume} {6}},\ \bibinfo {pages} {6400} (\bibinfo {year} {2015})}\BibitemShut {NoStop}%
\bibitem [{\citenamefont {Cepellotti}\ and\ \citenamefont {Marzari}(2017)}]{cepellotti2017transport}%
  \BibitemOpen
  \bibfield  {author} {\bibinfo {author} {\bibfnamefont {A.}~\bibnamefont {Cepellotti}}\ and\ \bibinfo {author} {\bibfnamefont {N.}~\bibnamefont {Marzari}},\ }\href@noop {} {\bibfield  {journal} {\bibinfo  {journal} {Physical Review Materials}\ }\textbf {\bibinfo {volume} {1}},\ \bibinfo {pages} {045406} (\bibinfo {year} {2017})}\BibitemShut {NoStop}%
\bibitem [{\citenamefont {Alvarez}(2018)}]{alvarez2018thermal}%
  \BibitemOpen
  \bibfield  {author} {\bibinfo {author} {\bibfnamefont {P.~T.}\ \bibnamefont {Alvarez}},\ }\href@noop {} {\emph {\bibinfo {title} {Thermal transport in semiconductors: first principles and phonon hydrodynamics}}}\ (\bibinfo  {publisher} {Springer},\ \bibinfo {year} {2018})\BibitemShut {NoStop}%
\bibitem [{\citenamefont {Chiloyan}\ \emph {et~al.}(2021)\citenamefont {Chiloyan}, \citenamefont {Huberman}, \citenamefont {Ding}, \citenamefont {Mendoza}, \citenamefont {Maznev}, \citenamefont {Nelson},\ and\ \citenamefont {Chen}}]{chiloyan2021green}%
  \BibitemOpen
  \bibfield  {author} {\bibinfo {author} {\bibfnamefont {V.}~\bibnamefont {Chiloyan}}, \bibinfo {author} {\bibfnamefont {S.}~\bibnamefont {Huberman}}, \bibinfo {author} {\bibfnamefont {Z.}~\bibnamefont {Ding}}, \bibinfo {author} {\bibfnamefont {J.}~\bibnamefont {Mendoza}}, \bibinfo {author} {\bibfnamefont {A.~A.}\ \bibnamefont {Maznev}}, \bibinfo {author} {\bibfnamefont {K.~A.}\ \bibnamefont {Nelson}}, \ and\ \bibinfo {author} {\bibfnamefont {G.}~\bibnamefont {Chen}},\ }\href@noop {} {\bibfield  {journal} {\bibinfo  {journal} {Physical Review B}\ }\textbf {\bibinfo {volume} {104}},\ \bibinfo {pages} {245424} (\bibinfo {year} {2021})}\BibitemShut {NoStop}%
\bibitem [{\citenamefont {Honarvar}\ \emph {et~al.}(2021)\citenamefont {Honarvar}, \citenamefont {Knobloch}, \citenamefont {Frazer}, \citenamefont {Abad}, \citenamefont {McBennett}, \citenamefont {Hussein}, \citenamefont {Kapteyn}, \citenamefont {Murnane},\ and\ \citenamefont {Hernandez-Charpak}}]{honarvar2021directional}%
  \BibitemOpen
  \bibfield  {author} {\bibinfo {author} {\bibfnamefont {H.}~\bibnamefont {Honarvar}}, \bibinfo {author} {\bibfnamefont {J.~L.}\ \bibnamefont {Knobloch}}, \bibinfo {author} {\bibfnamefont {T.~D.}\ \bibnamefont {Frazer}}, \bibinfo {author} {\bibfnamefont {B.}~\bibnamefont {Abad}}, \bibinfo {author} {\bibfnamefont {B.}~\bibnamefont {McBennett}}, \bibinfo {author} {\bibfnamefont {M.~I.}\ \bibnamefont {Hussein}}, \bibinfo {author} {\bibfnamefont {H.~C.}\ \bibnamefont {Kapteyn}}, \bibinfo {author} {\bibfnamefont {M.~M.}\ \bibnamefont {Murnane}}, \ and\ \bibinfo {author} {\bibfnamefont {J.~N.}\ \bibnamefont {Hernandez-Charpak}},\ }\href@noop {} {\bibfield  {journal} {\bibinfo  {journal} {Proceedings of the National Academy of Sciences}\ }\textbf {\bibinfo {volume} {118}},\ \bibinfo {pages} {e2109056118} (\bibinfo {year} {2021})}\BibitemShut {NoStop}%
\bibitem [{\citenamefont {Ward}\ and\ \citenamefont {Wilks}(1952)}]{ward1952iii}%
  \BibitemOpen
  \bibfield  {author} {\bibinfo {author} {\bibfnamefont {J.}~\bibnamefont {Ward}}\ and\ \bibinfo {author} {\bibfnamefont {J.}~\bibnamefont {Wilks}},\ }\href@noop {} {\bibfield  {journal} {\bibinfo  {journal} {The London, Edinburgh, and Dublin Philosophical Magazine and Journal of Science}\ }\textbf {\bibinfo {volume} {43}},\ \bibinfo {pages} {48} (\bibinfo {year} {1952})}\BibitemShut {NoStop}%
\bibitem [{\citenamefont {Chester}(1963)}]{chester1963second}%
  \BibitemOpen
  \bibfield  {author} {\bibinfo {author} {\bibfnamefont {M.}~\bibnamefont {Chester}},\ }\href@noop {} {\bibfield  {journal} {\bibinfo  {journal} {Physical Review}\ }\textbf {\bibinfo {volume} {131}},\ \bibinfo {pages} {2013} (\bibinfo {year} {1963})}\BibitemShut {NoStop}%
\bibitem [{\citenamefont {Sussmann}\ and\ \citenamefont {Thellung}(1963)}]{sussmann1963thermal}%
  \BibitemOpen
  \bibfield  {author} {\bibinfo {author} {\bibfnamefont {J.}~\bibnamefont {Sussmann}}\ and\ \bibinfo {author} {\bibfnamefont {A.}~\bibnamefont {Thellung}},\ }\href@noop {} {\bibfield  {journal} {\bibinfo  {journal} {Proceedings of the Physical Society}\ }\textbf {\bibinfo {volume} {81}},\ \bibinfo {pages} {1122} (\bibinfo {year} {1963})}\BibitemShut {NoStop}%
\bibitem [{\citenamefont {Enz}(1968)}]{enz1968one}%
  \BibitemOpen
  \bibfield  {author} {\bibinfo {author} {\bibfnamefont {C.~P.}\ \bibnamefont {Enz}},\ }\href@noop {} {\bibfield  {journal} {\bibinfo  {journal} {Annals of Physics}\ }\textbf {\bibinfo {volume} {46}},\ \bibinfo {pages} {114} (\bibinfo {year} {1968})}\BibitemShut {NoStop}%
\bibitem [{\citenamefont {Guyer}\ and\ \citenamefont {Krumhansl}(1966{\natexlab{b}})}]{guyer1966solution}%
  \BibitemOpen
  \bibfield  {author} {\bibinfo {author} {\bibfnamefont {R.~A.}\ \bibnamefont {Guyer}}\ and\ \bibinfo {author} {\bibfnamefont {J.}~\bibnamefont {Krumhansl}},\ }\href@noop {} {\bibfield  {journal} {\bibinfo  {journal} {Physical Review}\ }\textbf {\bibinfo {volume} {148}},\ \bibinfo {pages} {766} (\bibinfo {year} {1966}{\natexlab{b}})}\BibitemShut {NoStop}%
\bibitem [{\citenamefont {Beardo}\ \emph {et~al.}(2022)\citenamefont {Beardo}, \citenamefont {Alajlouni}, \citenamefont {Sendra}, \citenamefont {Bafaluy}, \citenamefont {Ziabari}, \citenamefont {Xuan}, \citenamefont {Camacho}, \citenamefont {Shakouri},\ and\ \citenamefont {Alvarez}}]{PhysRevB.105.165303}%
  \BibitemOpen
  \bibfield  {author} {\bibinfo {author} {\bibfnamefont {A.}~\bibnamefont {Beardo}}, \bibinfo {author} {\bibfnamefont {S.}~\bibnamefont {Alajlouni}}, \bibinfo {author} {\bibfnamefont {L.}~\bibnamefont {Sendra}}, \bibinfo {author} {\bibfnamefont {J.}~\bibnamefont {Bafaluy}}, \bibinfo {author} {\bibfnamefont {A.}~\bibnamefont {Ziabari}}, \bibinfo {author} {\bibfnamefont {Y.}~\bibnamefont {Xuan}}, \bibinfo {author} {\bibfnamefont {J.}~\bibnamefont {Camacho}}, \bibinfo {author} {\bibfnamefont {A.}~\bibnamefont {Shakouri}}, \ and\ \bibinfo {author} {\bibfnamefont {F.~X.}\ \bibnamefont {Alvarez}},\ }\href {\doibase 10.1103/PhysRevB.105.165303} {\bibfield  {journal} {\bibinfo  {journal} {Phys. Rev. B}\ }\textbf {\bibinfo {volume} {105}},\ \bibinfo {pages} {165303} (\bibinfo {year} {2022})}\BibitemShut {NoStop}%
\bibitem [{\citenamefont {Zeng}\ and\ \citenamefont {Dong}(2019)}]{zeng_and_dong_2019_vibFPE}%
  \BibitemOpen
  \bibfield  {author} {\bibinfo {author} {\bibfnamefont {Y.}~\bibnamefont {Zeng}}\ and\ \bibinfo {author} {\bibfnamefont {J.}~\bibnamefont {Dong}},\ }\href@noop {} {\bibfield  {journal} {\bibinfo  {journal} {Physical Review B}\ }\textbf {\bibinfo {volume} {99}},\ \bibinfo {pages} {014306} (\bibinfo {year} {2019})}\BibitemShut {NoStop}%
\bibitem [{\citenamefont {Hua}\ and\ \citenamefont {Lindsay}(2020)}]{hua2020space}%
  \BibitemOpen
  \bibfield  {author} {\bibinfo {author} {\bibfnamefont {C.}~\bibnamefont {Hua}}\ and\ \bibinfo {author} {\bibfnamefont {L.}~\bibnamefont {Lindsay}},\ }\href@noop {} {\bibfield  {journal} {\bibinfo  {journal} {Physical Review B}\ }\textbf {\bibinfo {volume} {102}},\ \bibinfo {pages} {104310} (\bibinfo {year} {2020})}\BibitemShut {NoStop}%
\bibitem [{\citenamefont {Gurtin}\ and\ \citenamefont {Pipkin}(1968)}]{gurtin_and_pipkin_1968_kern}%
  \BibitemOpen
  \bibfield  {author} {\bibinfo {author} {\bibfnamefont {M.~E.}\ \bibnamefont {Gurtin}}\ and\ \bibinfo {author} {\bibfnamefont {A.~C.}\ \bibnamefont {Pipkin}},\ }\href@noop {} {\bibfield  {journal} {\bibinfo  {journal} {Archive for Rational Mechanics and Analysis}\ }\textbf {\bibinfo {volume} {31}},\ \bibinfo {pages} {113} (\bibinfo {year} {1968})}\BibitemShut {NoStop}%
\bibitem [{\citenamefont {Nunziato}(1971)}]{nunziato_1971_memory}%
  \BibitemOpen
  \bibfield  {author} {\bibinfo {author} {\bibfnamefont {J.~W.}\ \bibnamefont {Nunziato}},\ }\href@noop {} {\bibfield  {journal} {\bibinfo  {journal} {Quarterly of Applied Mathematics}\ }\textbf {\bibinfo {volume} {29}},\ \bibinfo {pages} {187} (\bibinfo {year} {1971})}\BibitemShut {NoStop}%
\bibitem [{\citenamefont {Alvarez}\ and\ \citenamefont {Jou}(2007)}]{alvarez2007memory}%
  \BibitemOpen
  \bibfield  {author} {\bibinfo {author} {\bibfnamefont {F.}~\bibnamefont {Alvarez}}\ and\ \bibinfo {author} {\bibfnamefont {D.}~\bibnamefont {Jou}},\ }\href@noop {} {\bibfield  {journal} {\bibinfo  {journal} {Applied physics letters}\ }\textbf {\bibinfo {volume} {90}} (\bibinfo {year} {2007})}\BibitemShut {NoStop}%
\bibitem [{\citenamefont {Zwanzig}(1961)}]{zwanzig1961memory}%
  \BibitemOpen
  \bibfield  {author} {\bibinfo {author} {\bibfnamefont {R.}~\bibnamefont {Zwanzig}},\ }\href@noop {} {\bibfield  {journal} {\bibinfo  {journal} {Physical Review}\ }\textbf {\bibinfo {volume} {124}},\ \bibinfo {pages} {983} (\bibinfo {year} {1961})}\BibitemShut {NoStop}%
\bibitem [{\citenamefont {Zwanzig}(1965{\natexlab{b}})}]{zwanzig1965frequency}%
  \BibitemOpen
  \bibfield  {author} {\bibinfo {author} {\bibfnamefont {R.}~\bibnamefont {Zwanzig}},\ }\href@noop {} {\bibfield  {journal} {\bibinfo  {journal} {The Journal of Chemical Physics}\ }\textbf {\bibinfo {volume} {43}},\ \bibinfo {pages} {714} (\bibinfo {year} {1965}{\natexlab{b}})}\BibitemShut {NoStop}%
\bibitem [{\citenamefont {Shanks}\ \emph {et~al.}(1963)\citenamefont {Shanks}, \citenamefont {Maycock}, \citenamefont {Sidles},\ and\ \citenamefont {Danielson}}]{PhysRev.130.1743}%
  \BibitemOpen
  \bibfield  {author} {\bibinfo {author} {\bibfnamefont {H.~R.}\ \bibnamefont {Shanks}}, \bibinfo {author} {\bibfnamefont {P.~D.}\ \bibnamefont {Maycock}}, \bibinfo {author} {\bibfnamefont {P.~H.}\ \bibnamefont {Sidles}}, \ and\ \bibinfo {author} {\bibfnamefont {G.~C.}\ \bibnamefont {Danielson}},\ }\href {\doibase 10.1103/PhysRev.130.1743} {\bibfield  {journal} {\bibinfo  {journal} {Phys. Rev.}\ }\textbf {\bibinfo {volume} {130}},\ \bibinfo {pages} {1743} (\bibinfo {year} {1963})}\BibitemShut {NoStop}%
\bibitem [{\citenamefont {Glassbrenner}\ and\ \citenamefont {Slack}(1964)}]{PhysRev.134.A1058}%
  \BibitemOpen
  \bibfield  {author} {\bibinfo {author} {\bibfnamefont {C.~J.}\ \bibnamefont {Glassbrenner}}\ and\ \bibinfo {author} {\bibfnamefont {G.~A.}\ \bibnamefont {Slack}},\ }\href {\doibase 10.1103/PhysRev.134.A1058} {\bibfield  {journal} {\bibinfo  {journal} {Phys. Rev.}\ }\textbf {\bibinfo {volume} {134}},\ \bibinfo {pages} {A1058} (\bibinfo {year} {1964})}\BibitemShut {NoStop}%
\bibitem [{\citenamefont {Fulkerson}\ \emph {et~al.}(1968)\citenamefont {Fulkerson}, \citenamefont {Moore}, \citenamefont {Williams}, \citenamefont {Graves},\ and\ \citenamefont {McElroy}}]{PhysRev.167.765}%
  \BibitemOpen
  \bibfield  {author} {\bibinfo {author} {\bibfnamefont {W.}~\bibnamefont {Fulkerson}}, \bibinfo {author} {\bibfnamefont {J.~P.}\ \bibnamefont {Moore}}, \bibinfo {author} {\bibfnamefont {R.~K.}\ \bibnamefont {Williams}}, \bibinfo {author} {\bibfnamefont {R.~S.}\ \bibnamefont {Graves}}, \ and\ \bibinfo {author} {\bibfnamefont {D.~L.}\ \bibnamefont {McElroy}},\ }\href {\doibase 10.1103/PhysRev.167.765} {\bibfield  {journal} {\bibinfo  {journal} {Phys. Rev.}\ }\textbf {\bibinfo {volume} {167}},\ \bibinfo {pages} {765} (\bibinfo {year} {1968})}\BibitemShut {NoStop}%
\bibitem [{\citenamefont {Vega-Flick}\ \emph {et~al.}(2016)\citenamefont {Vega-Flick}, \citenamefont {Duncan}, \citenamefont {Eliason}, \citenamefont {Cuffe}, \citenamefont {Johnson}, \citenamefont {Peraud}, \citenamefont {Zeng}, \citenamefont {Lu}, \citenamefont {Maznev}, \citenamefont {Wang} \emph {et~al.}}]{vega2016thermal}%
  \BibitemOpen
  \bibfield  {author} {\bibinfo {author} {\bibfnamefont {A.}~\bibnamefont {Vega-Flick}}, \bibinfo {author} {\bibfnamefont {R.~A.}\ \bibnamefont {Duncan}}, \bibinfo {author} {\bibfnamefont {J.~K.}\ \bibnamefont {Eliason}}, \bibinfo {author} {\bibfnamefont {J.}~\bibnamefont {Cuffe}}, \bibinfo {author} {\bibfnamefont {J.~A.}\ \bibnamefont {Johnson}}, \bibinfo {author} {\bibfnamefont {J.-P.}\ \bibnamefont {Peraud}}, \bibinfo {author} {\bibfnamefont {L.}~\bibnamefont {Zeng}}, \bibinfo {author} {\bibfnamefont {Z.}~\bibnamefont {Lu}}, \bibinfo {author} {\bibfnamefont {A.~A.}\ \bibnamefont {Maznev}}, \bibinfo {author} {\bibfnamefont {E.~N.}\ \bibnamefont {Wang}},  \emph {et~al.},\ }\href@noop {} {\bibfield  {journal} {\bibinfo  {journal} {AIP advances}\ }\textbf {\bibinfo {volume} {6}} (\bibinfo {year} {2016})}\BibitemShut {NoStop}%
\bibitem [{\citenamefont {Cahill}(2004)}]{cahill2004analysis}%
  \BibitemOpen
  \bibfield  {author} {\bibinfo {author} {\bibfnamefont {D.~G.}\ \bibnamefont {Cahill}},\ }\href@noop {} {\bibfield  {journal} {\bibinfo  {journal} {Review of scientific instruments}\ }\textbf {\bibinfo {volume} {75}},\ \bibinfo {pages} {5119} (\bibinfo {year} {2004})}\BibitemShut {NoStop}%
\bibitem [{\citenamefont {Kresse}\ and\ \citenamefont {Furthm{\"u}ller}(1996)}]{Kresse1996}%
  \BibitemOpen
  \bibfield  {author} {\bibinfo {author} {\bibfnamefont {G.}~\bibnamefont {Kresse}}\ and\ \bibinfo {author} {\bibfnamefont {J.}~\bibnamefont {Furthm{\"u}ller}},\ }\href@noop {} {\bibfield  {journal} {\bibinfo  {journal} {Physical Review B}\ }\textbf {\bibinfo {volume} {54}},\ \bibinfo {pages} {11169} (\bibinfo {year} {1996})}\BibitemShut {NoStop}%
\end{thebibliography}%
\end{document}